\documentclass[pdflatex,sn-mathphys-num]{sn-jnl}

\usepackage{amsmath,amssymb,amsfonts,mathrsfs}
\usepackage{graphicx}
\usepackage{subcaption}
\usepackage{booktabs}
\usepackage{array}
\usepackage{multirow}
\usepackage{xcolor}
\usepackage{adjustbox}
\usepackage{tabularx}
\usepackage{makecell}
\usepackage{threeparttable}

\usepackage{tikz}
\usetikzlibrary{arrows.meta,positioning,calc,shapes.geometric,fit,shadows.blur}
\usepackage{pgfplots}
\pgfplotsset{compat=1.18}
\usepgfplotslibrary{groupplots,fillbetween}

\usepackage[ruled,vlined,linesnumbered]{algorithm2e}
\usepackage{listings}
\usepackage[title]{appendix}
\usepackage{manyfoot}
\usepackage{placeins}
\PassOptionsToPackage{hidelinks}{hyperref}
\usepackage{hyperref}

\newcommand{\revb}[1]{#1}

\newif\ifdraftflags \draftflagstrue

\theoremstyle{thmstyleone}

\theoremstyle{thmstyletwo}

\theoremstyle{thmstylethree}

\begin{document}

\title[Article Title]{Multiphysics Modeling of Thermo-Viscoelastic Damage
in Functionally Graded Abradable Coatings with
Probabilistic Geometric Tolerance Analysis}

\author[1]{\fnm{Amjad} \sur{El-Mellouhi}}\email{amjad.el-mellouhi@polymtl.ca}
\author[2]{\fnm{Yassine} \sur{Adjal}}\email{yassine.adjal@univusto.dz}
\author[3]{\fnm{Khaled} \sur{Dhibi}}\email{kdhibi@hbku.edu.qa}
\author*[3]{\fnm{Fedwa} \sur{El-Mellouhi}}\email{felmellouhi@hbku.edu.qa}

\affil[1]{\orgname{Polytechnique Montréal}, \city{Montréal}, \state{Québec}, \country{Canada}}
\affil[2]{\orgdiv{Department of Maritime Engineering}, \orgname{University of Sciences and Technology Mohamed Boudiaf}, \orgaddress{\street{BP 1505 El M'naouer}, \city{Oran}, \country{Algeria}}}
\affil[3]{\orgdiv{College of Science and Engineering}, \orgname{Hamad Bin Khalifa University}, \city{Doha}, \postcode{PoBox 34110}, \country{Qatar}}

\abstract{
\textcolor{black}{
In aircraft engines, functionally graded abradable coatings are used to control
blade-tip clearance, but their durability is governed by effects that are often
treated separately in existing models, including temperature-dependent
viscoelastic softening, progressive damage, deposition-induced microstructural
modulation, and geometric tolerances. Models that consider these effects
independently may underestimate damage initiation because smooth material
gradients evaluated at nominal geometry cannot capture the local stress
concentrations produced by microstructural modulation and geometric variability.
This study integrates these effects within a unified
multiphysics--probabilistic framework. The solved domain is a local
through-thickness coating column driven by prescribed strain and temperature
histories that include thermal eigenstrain and coating--substrate expansion
mismatch.  The results show that
periodic property modulation increases end-of-cycle damage relative to the
monotonic gradient and concentrates the maximum stress within a localized
modulation crest. Propagating symmetric geometric tolerances through the coupled
solver produces a strongly right-skewed damage distribution and a higher
classification-level exceedance probability for the modulated gradient than for
the monotonic gradient. This exceedance behavior cannot be obtained from a
deterministic analysis performed at nominal geometry. The exceedance probability
is also evaluated as a continuous function of the classification level to assess
the sensitivity of the comparison to the selected screening criterion. The
implementation is verified through spatial and temporal refinement,
closed-form unit tests, nested Monte Carlo convergence with Wilson confidence
intervals, and a hold-out-validated polynomial response surface. Numerical
verification is clearly distinguished from experimental validation, and the
simulated case is presented as a generic benchmark rather than an identified
material model. The framework links deposition parameters and tolerance bands
to damage-exceedance risk, supporting reliability-oriented screening of graded
and multilayer coating systems.
}}

\keywords{Abradable coatings, Functionally graded materials,
Thermo-viscoelasticity, Damage evolution, Geometric tolerances,
Monte Carlo simulation, Reliability analysis, Turbine labyrinth seals}

\maketitle

\section{Introduction}
\label{sec:intro}

Labyrinth seals equipped with abradable coatings are critical elements in
modern aircraft gas turbines, ensuring effective dynamic sealing while limiting
wear of rotating
components~\cite{pathak2022ysz,baillieu2024insitu,bertuol2025alsi}.
\revb{The rubbing behaviour of labyrinth seals against abradable inlet liners has
been characterized experimentally in dedicated rub-rig campaigns, notably in the
systematic study of Pychynski on honeycomb inlet
liners~\cite{pychynski2016labyrinth}, which documents the coupling between
incursion kinematics, contact forces and frictional heating that motivates the
equivalent prescribed histories adopted here.} Because
sealing performance directly influences leakage losses, clearance control, and
thermal margins, the thermo-mechanical response of abradable coatings has a
direct impact on engine efficiency, durability, and operational
safety~\cite{bertuol2025alsi,schapery1969nonlinear,simo1996schapery}. Continued
increases in operating temperatures, reduced functional clearances, and longer
service intervals have progressively pushed conventional coating architectures
to their performance limits~\cite{brinson2008polymer,lemaitre2005engineering}.

From a mechanics standpoint, abradable coatings exhibit a coupled response
driven by temperature-dependent thermo-viscoelasticity and progressive
degradation under cyclic thermo-mechanical loading. This coupling induces
evolving stress redistributions and localized damage accumulation that govern
early crack initiation and durability. However, many existing numerical
approaches remain based on linear viscoelastic formulations and/or global
damage laws~\cite{kachanov1986damage,lakes2009viscoelasticmaterials,%
wineman2000mechanical}, which can be insufficient to resolve localized stress
concentrations and the associated spatial evolution of damage, especially under
strongly heterogeneous material
fields~\cite{schapery1969nonlinear,simo1996schapery,lemaitre2005engineering}.
The viscoelastic memory effects and their interaction with progressive stiffness
degradation require a tightly coupled formulation that resolves both temporal
stress relaxation and spatial damage
localization~\cite{ferry1980viscoelastic,kachanov1986damage}.

Functionally graded materials (FGMs) have emerged as a promising design
strategy, as their continuous through-thickness variation of mechanical and
thermal properties can reduce discontinuities and mitigate stress
concentrations~\cite{wang2025ceramic,suresh1998fundamentals,miyamoto1999fgm}.
Prior investigations have shown that graded architectures may improve damage
tolerance and reduce peak stresses compared with homogeneous or sharply layered
coatings~\cite{wang2025ceramic,suresh1998fundamentals}. Nevertheless, most
modeling studies assume idealized monotonic gradients and overlook periodic
microstructural modulations induced by deposition
processes~\cite{huang2024review,lokachari2024columnar}. Such deposition-driven
heterogeneities can create local stiffness contrasts that amplify stresses and
trigger early damage initiation, yet they are rarely incorporated explicitly in
continuum-scale analyses~\cite{haldar2000probability,yildirim2011periodic,%
chen2021modulation}.

In addition to material heterogeneity, manufacturing and assembly processes
inevitably introduce geometric tolerances in labyrinth
seals~\cite{schapery1969nonlinear,ferry1980viscoelastic,huang2024review,%
haldar2000probability,ding2021probabilistic}. Even small dimensional deviations
can lead to significant variability in local contact conditions and
thermo-mechanical fields, thereby affecting stress hotspots and damage
accumulation. Although probabilistic methods are well established in structural
reliability analysis~\cite{wang2025ceramic,ditlevsen1996reliability,%
melchers1999structural}, their integration into thermo-viscoelastic damage
modeling of abradable coatings remains limited, despite the fact that rare
worst-case geometries may dominate reliability and maintenance risk.

Addressing these gaps requires more than combining established ingredients; the
key advance lies in their unified treatment. When periodic property modulation
and geometric tolerances are resolved simultaneously within a single
thermo-viscoelastic damage solver, they interact in ways that neither analysis
can capture independently. In particular, the spatially varying property field
governs local stress concentrations that drive damage, while geometric
tolerances perturb the prescribed equivalent mechanical excitation. Their
coupled propagation through the damage solver produces nonlinear stress and damage
distributions that cannot be obtained by treating the material architecture and the
geometric deviations
separately~\cite{haldar2000probability,ding2021probabilistic,%
ditlevsen1996reliability}. As a result, probabilistic stress and damage
predictions reflect the nonlinear interaction between the deterministic modulated
property field and the sampled geometric deviations. Embedding these effects within one computational pipeline enables
manufacturing-informed design decisions: functional-gradient exponents,
deposition-modulation amplitudes, and dimensional-tolerance bands can be
screened against quantified damage-exceedance probabilities, thereby linking
manufacturing parameters directly to structural reliability metrics in a way
that prior decoupled approaches do not
provide~\cite{huang2024review,haldar2000probability,melchers1999structural}.

The individual modeling ingredients, Prony-series viscoelasticity, scalar damage
mechanics, functional grading, and Monte Carlo reliability analysis, are each
well established in the
literature~\cite{schapery1969nonlinear,kachanov1986damage,%
suresh1998fundamentals,ditlevsen1996reliability}. The novelty of the present
contribution lies in three aspects of their integration: (i)~periodic
microstructural modulation is embedded directly in the constitutive property
field rather than introduced as an external perturbation, so its influence on
damage is captured self-consistently through the coupled
solver~\cite{yildirim2011periodic,chen2021modulation}; (ii)~geometric tolerances
are propagated through the full multiphysics chain rather than through a reduced
surrogate, preserving the fidelity of local stress and damage predictions under
dimensional variability~\cite{ding2021probabilistic,melchers1999structural};
and (iii)~the resulting probabilistic outputs, including exceedance
probabilities, sensitivity indices, and worst-case configurations, are expressed
in forms that can be used directly for tolerance specification and maintenance
planning~\cite{ditlevsen1996reliability,sudret2008pce}. Although demonstrated
here for abradable coatings in turbine labyrinth seals, the mathematical
structure is sufficiently general to apply to other graded or layered systems
subjected to cyclic thermo-mechanical loading in which deposition-induced
heterogeneity and dimensional variability jointly govern performance, including
thermal barrier coatings, environmental barrier coatings for ceramic-matrix
composites, and functionally graded structural components in aerospace and
energy applications~\cite{miyamoto1999fgm,li2021stochastic}.

This study aligns closely with themes central to advanced materials and
manufacturing research. The framework follows a quantitative
process--structure--property--performance chain: deposition-process parameters
govern microstructural modulation (structure), which in turn shapes local
stiffness and stress fields (property), ultimately influencing damage
accumulation and service reliability (performance). Manufacturing variability is
introduced explicitly through probabilistic tolerance
propagation~\cite{haldar2000probability,sudret2008pce}, yielding reliability
metrics that can support tolerance specification, quality-control objectives,
and condition-based maintenance
planning~\cite{ding2021probabilistic,dhibi2026physics}. By linking materials
response, manufacturing-induced variability, and reliability-oriented design
within a unified framework, the present work responds to the growing need for
process-aware and uncertainty-quantified modeling tools in advanced coating
manufacturing~\cite{miyamoto1999fgm,huang2024review}.

To address these gaps, this work proposes a unified multiphysics modeling
framework that couples: (i)~temperature-dependent thermo-viscoelasticity using a
Prony-series representation~\cite{bertuol2025alsi,schapery1969nonlinear,%
simo1996schapery,ferry1980viscoelastic}; (ii)~a stress-driven scalar damage
formulation governed by a critical stress
threshold~\cite{lemaitre2005engineering,kachanov1986damage}; (iii)~a periodically
modulated functional gradient to represent deposition-induced microstructural
heterogeneity~\cite{huang2024review,yildirim2011periodic,chen2021modulation};
and (iv)~probabilistic propagation of geometric tolerances using Monte Carlo
simulations~\cite{haldar2000probability,ding2021probabilistic,%
ditlevsen1996reliability}. This integrated approach enables prediction of both
the nominal response and the variability of localized stress and damage metrics,
thereby supporting reliability-oriented assessment and design guidance for
next-generation abradable coatings.

{\color{black}%

Because the credibility of a multiphysics framework depends on what it does
not claim as much as on what it computes, the status of every ingredient is
declared here and maintained consistently throughout the paper.

The physical system of interest is a labyrinth-seal stage in which relative
radial motion between the sealing teeth and the coated counter-face may close
the running clearance and produce a localized rubbing region. In that physical
system, normal pressure, tangential traction, frictional heating, and material
removal are all real phenomena. The solved domain in this work, however, is a
local through-thickness coating column divided into perfectly bonded
homogenized subdomains and driven by prescribed temperature and equivalent
normal-strain histories, denoted by $T(z,t)$ and
$\varepsilon^{\mathrm{app}}(t)$, respectively. The reduced solver contains no
tangential degree of freedom, no sliding interface, no contact-pressure
calculation, no Coulomb friction coefficient, and no material-removal
kinetics. Consequently, no wear depth, incursion depth, or service life is
reported as a computed quantity anywhere in this manuscript.
Table~\ref{tab:scope} states this modeling boundary for each quantity, and
Fig.~\ref{fig:config} presents the physical system and the solved domain
separately.

The simulated case is a declared generic benchmark. Its parameter set is
literature-consistent for thermally sprayed AlSi-based abradable systems but is
not presented as an identified material model for a specific coating. Damage
parameters are anchored on published elevated-temperature data
(Section~\ref{subsec:calib}); the extension of the benchmark cycle beyond the
calibration temperature is a constitutive model-based extrapolation, evaluated
as a severe numerical case rather than as experimental validation under that
condition. Throughout the paper, numerical verification, including refinement
studies, unit tests, sampling convergence, and surrogate cross-checking, is
reported and explicitly distinguished from experimental validation, which
remains future work.
}

The remainder of this paper is organized as follows.
Section~\ref{sec:framework} presents the proposed framework.
Section~\ref{sec:theoretical_modeling} presents the theoretical formulation of
the coupled thermo-viscoelastic--damage model with graded and periodically
modulated properties. Section~\ref{sec:numerical_methodology} describes the
numerical implementation, the verification studies, and the probabilistic
tolerance framework. Section~\ref{sec:results_discussion} discusses
deterministic trends and probabilistic response distributions.
Section~\ref{sec:implications_limits_prospects} summarizes practical
implications, limitations, and perspectives for future developments.
Section~\ref{sec:conclusions} concludes the paper.
\section{Proposed Framework}
\label{sec:framework}

\begin{figure}[!t]
\centering
\begin{adjustbox}{max width=\linewidth}
\begin{tikzpicture}[
  font=\footnotesize, >=Latex,
  blk/.style   = {draw=gray!70, rounded corners=2pt, align=left,
                  inner sep=4pt, text width=36mm, font=\footnotesize},
  grp/.style   = {draw=blue!45, dashed, rounded corners=4pt, inner sep=7pt},
  ar/.style    = {-{Latex[length=4pt]}, gray!70, line width=0.7pt}
]
\node[blk] (in)  {\textbf{Inputs}\\
  -- Nominal seal geometry\\ -- Prescribed thermal $+$ mechanical histories\\
  -- Baseline FGM profile};
\node[blk, below=5mm of in] (pf) {\textbf{Property-field construction}\\
  -- Functional gradient (FGM)\\ -- Periodic modulation\\
  -- Temperature dependence};
\node[blk, below=5mm of pf] (sv) {\textbf{Coupled solver}\\
  -- Thermo-viscoelasticity\\ -- Thermal eigenstrain\\ -- Stress-driven damage};
\node[blk, below=5mm of sv] (dt) {\textbf{Deterministic outputs}\\
  -- Stress/strain fields\\ -- Damage evolution maps\\ -- Hot-spot localization};
\node[grp, fit=(in)(pf)(sv)(dt), label={[font=\scriptsize\bfseries, blue!60!black]above:Deterministic multiphysics core}] (core) {};

\node[blk, right=26mm of in] (tm) {\textbf{Tolerance model}\\
  -- Clearance/offset deviations\\ -- Manufacturing bounds\\ -- Assembly limits};
\node[blk, below=5mm of tm] (mc) {\textbf{Monte Carlo sampling}\\
  -- Sample geometries $\delta\mathbf{u}$\\ -- Generate $N$ realizations};
\node[blk, below=5mm of mc] (rn) {\textbf{Run solver per sample}\\
  -- Evaluate multiphysics core\\ -- Parallel-ready workflow};
\node[blk, below=5mm of rn] (ro) {\textbf{Reliability outputs}\\
  -- Variability envelopes\\ -- Exceedance probability\\ -- Worst-case configurations};
\node[grp, fit=(tm)(mc)(rn)(ro), label={[font=\scriptsize\bfseries, blue!60!black]above:Probabilistic tolerance propagation}] (prob) {};

\draw[ar] (in) -- (pf);  \draw[ar] (pf) -- (sv);  \draw[ar] (sv) -- (dt);
\draw[ar] (tm) -- (mc);  \draw[ar] (mc) -- (rn);  \draw[ar] (rn) -- (ro);
\draw[ar] (core.east|-mc) -- (mc.west);
\draw[ar] (rn.west) -- (rn.west-|core.east);
\end{tikzpicture}
\end{adjustbox}
\caption{Block diagram of the proposed multiphysics--probabilistic framework. A
deterministic multiphysics core (property-field construction $+$ coupled
thermo-viscoelastic--damage solver) is wrapped by a probabilistic tolerance
propagation layer (Monte Carlo sampling of geometric deviations) to produce both
nominal predictions and reliability metrics.}
\label{fig:framework}
\end{figure}
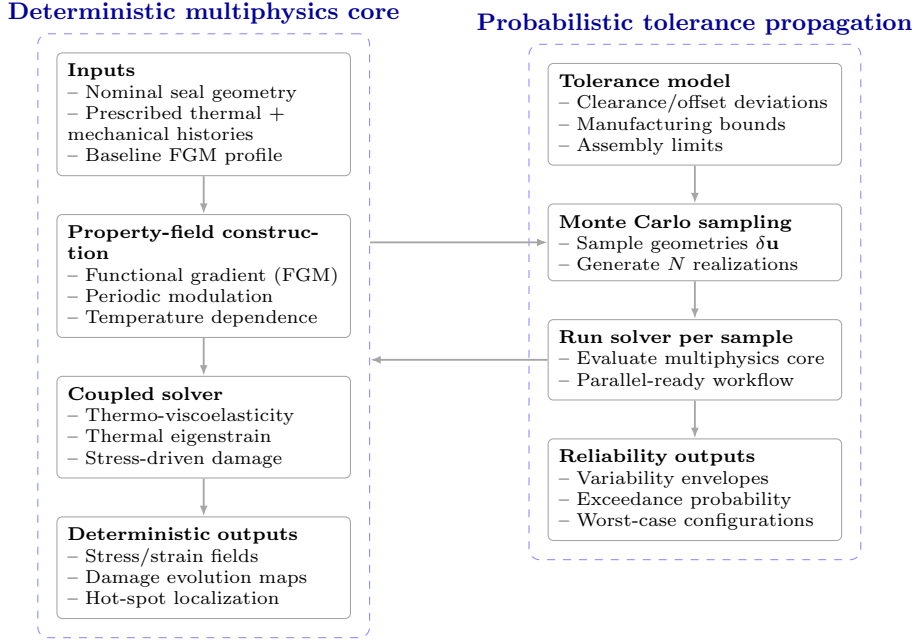

Fig.~\ref{fig:framework} summarizes the proposed multiphysics--probabilistic
framework for predicting the thermo-mechanical response and reliability of
labyrinth-seal abradable coatings under cyclic thermo-mechanical loading. The
purpose of the framework is not only to compute nominal stress and damage
fields, but also to propagate manufacturing-induced variability through the full
model so that design decisions can be informed by both expected response and
reliability risk. In this sense, the proposed approach acts as an integrated
design-assessment and reliability-evaluation tool rather than as a stand-alone
constitutive or numerical model.

The framework is designed to capture four coupled durability drivers:
(i)~temperature-dependent thermo-viscoelastic deformation described through a
Prony-series, or generalized Maxwell,
representation~\cite{schapery1969nonlinear,simo1996schapery,%
ferry1980viscoelastic}; (ii)~progressive spatial damage evolution governed by a
stress-driven scalar damage law~\cite{lemaitre2005engineering,%
kachanov1986damage}; (iii)~deposition-induced periodic microstructural
heterogeneity superimposed on the functional
gradient~\cite{huang2024review,lokachari2024columnar}; and (iv)~geometric
tolerance variability that affects the local
\revb{prescribed excitation and the resulting}
stress hot spots, propagated using Monte Carlo
simulation~\cite{haldar2000probability,ditlevsen1996reliability}. Their
integration enables direct mapping from manufacturing choices and dimensional
dispersion to stress localization, damage accumulation, and exceedance risk,
thereby strengthening the process--structure--property--performance
interpretation of the framework.

\begin{itemize}
  \item \textbf{Stage 0: Inputs and problem definition.}
  The workflow starts from three input groups: (i)~nominal configuration,
  including seal geometry, interfaces, and boundary definitions
  \revb{(Section~\ref{subsec:config}, Fig.~\ref{fig:config})}; (ii)~loading
  history, including the thermal cycle and the corresponding
  \revb{prescribed equivalent mechanical excitation
  (Section~\ref{subsec:loading}, Fig.~\ref{fig:cycle})}; and (iii)~material
  architecture, represented by the baseline through-thickness grading law.
  These inputs define the initial process and design space from which both
  deterministic response and variability-aware reliability assessment are
  derived.

  \item \textbf{Stage 1: Property-field construction.}
  A material-field generator combines: (i)~the baseline functional gradient,
  (ii)~periodic modulation representing deposition-induced microstructural
  oscillations~\cite{huang2024review,lokachari2024columnar}, and
  (iii)~temperature dependence of viscoelastic
  parameters~\cite{schapery1969nonlinear,ferry1980viscoelastic}. The output is a
  temperature-aware spatial property map passed to the deterministic solver.
  This stage explicitly links manufacturing-induced heterogeneity to the local
  constitutive description of the coating, allowing process-driven
  microstructural modulation to influence local stiffness, stress
  redistribution, and subsequent damage evolution.

  \item \textbf{Stage 2: Deterministic multiphysics core.}
  The deterministic core solves the coupled problem over the prescribed cycle:
  thermo-viscoelastic response via a Prony-series
  representation~\cite{schapery1969nonlinear,simo1996schapery,%
  ferry1980viscoelastic}, \revb{including the thermal eigenstrain and the
  coating--substrate expansion mismatch of
  Section~\ref{subsec:viscoelastic},} and progressive stiffness degradation via
  a stress-driven scalar damage model~\cite{lemaitre2005engineering,%
  kachanov1986damage}. As a result, the framework captures how local stiffness
  evolution, stress redistribution, and damage localization interact over time
  under coupled thermo-mechanical loading. This stage provides the nominal
  stress--damage response against which the effect of manufacturing variability
  can be assessed.

  \item \textbf{Stage 3: Probabilistic tolerance propagation.}
  Geometric tolerances are modeled probabilistically and sampled to generate $N$
  realizations~\cite{haldar2000probability}. The deterministic core is evaluated
  for each realization, with the simulations parallelizable across samples,
  producing response distributions and reliability
  statistics~\cite{ditlevsen1996reliability}. This step extends the analysis
  beyond nominal predictions by quantifying how manufacturing and assembly
  dispersion affect durability risk. Because each sampled geometry is propagated
  through the full multiphysics solver, the resulting reliability metrics
  preserve the coupled influence of material heterogeneity, boundary
  perturbations, stress localization, and damage evolution.
  \revb{A polynomial response surface is additionally trained and independently
  validated against the same solver, and is used as a cross-check and screening
  tool rather than as a replacement for the reference estimator
  (Section~\ref{subsec:prob}).}

  \item \textbf{Stage 4: Outputs.}
  The framework delivers deterministic outputs, including stress/strain fields,
  damage maps, effective stiffness evolution, and hot-spot localization,
  together with probabilistic outputs, including variability envelopes,
  exceedance probabilities, sensitivity indices, and worst-case
  configurations~\cite{haldar2000probability,ditlevsen1996reliability}.
  Collectively, these outputs support manufacturing-informed design, tolerance
  control, coating-architecture optimization, and reliability-oriented
  decision-making for abradable coating systems. In particular, they provide
  quantitative links between process parameters, geometric tolerances, local
  damage risk, and maintenance-relevant reliability metrics.
\end{itemize}

\section{Theoretical Modeling}
\label{sec:theoretical_modeling}

{\color{black}%
\subsection{Material system, physical configuration, and model scope}
\label{subsec:config}

\subsubsection{Material system and baseline properties}
\label{subsec:material}
The reference material family is a thermally sprayed AlSi-based abradable
system, in which an aluminium--silicon matrix is combined with a soft
dislocator phase such as hexagonal boron nitride (hBN), representative of
abradable seals used in the compressor sections of
aero-engines~\cite{bertuol2025alsi,faraoun2006abradable}. As-sprayed coatings of
this family are strongly heterogeneous, with substantial total porosity
distributed quasi-periodically through the thickness as a result of the
pass-by-pass deposition sequence~\cite{faraoun2006abradable,%
lokachari2024columnar}. Table~\ref{tab:baseline} summarizes the baseline
material families, indicative porosity levels, and the corresponding sources.

Porosity is not represented by meshing individual pores. It is introduced
through an effective homogenized description: the porosity level associated with
each baseline material is incorporated into the effective properties assigned to
the homogenized subdomains, while the periodic modulation of
Eq.~\eqref{eq:fgm_modulus} represents the prescribed spatial variation of those
effective properties through the coating thickness. This treatment is consistent
with the homogeneous multilayer representation used for functionally graded
coatings by Liu et al.~\cite{liu2012thermoelastic}. Two consequences follow and
are stated as limitations in Section~\ref{sec:implications_limits_prospects}:
pore-scale stress concentrations are not resolved, and the thermal-barrier role
of porosity does not enter the solution because the temperature history is
prescribed per subdomain rather than obtained from a heat-conduction problem.

\begin{table}[p]
\centering
\caption{Summary of key model parameters, \revb{numerical values,} and
uncertainty definitions (units and origin), covering functionally graded material
modeling, damage mechanics, and probabilistic reliability analysis
\cite{lemaitre2005engineering,suresh1998fundamentals,yildirim2011periodic,%
ditlevsen1996reliability}. \revb{Every parameter is listed with its
numerical value, its origin, and its declared status. Benchmark denotes a declared
numerical choice for the generic verification case; identification required
denotes a quantity that a material-specific study would have to measure.}}
\label{tab:param_summary}

\vspace{-1mm}

\begingroup
\fontsize{4.8pt}{5.3pt}\selectfont
\renewcommand{\arraystretch}{0.80}
\setlength{\tabcolsep}{1.0pt}
\setlength{\aboverulesep}{0.15ex}
\setlength{\belowrulesep}{0.15ex}

\begin{tabular}{
>{\raggedright\arraybackslash}
 p{\dimexpr 0.1518\linewidth-2\tabcolsep\relax}
>{\raggedright\arraybackslash}
 p{\dimexpr 0.1128\linewidth-2\tabcolsep\relax}
>{\raggedright\arraybackslash}
 p{\dimexpr 0.2374\linewidth-2\tabcolsep\relax}
>{\raggedright\arraybackslash}
 p{\dimexpr 0.0661\linewidth-2\tabcolsep\relax}
>{\raggedright\arraybackslash}
 p{\dimexpr 0.1790\linewidth-2\tabcolsep\relax}
>{\raggedright\arraybackslash}
 p{\dimexpr 0.2529\linewidth-2\tabcolsep\relax}
}
\toprule
\textbf{Group}
&
\textbf{Symbol}
&
\textbf{Meaning}
&
\textbf{Unit}
&
\revb{\textbf{Value}}
&
\textbf{Origin / note}
\\
\midrule

\multicolumn{6}{l}{\textbf{Geometry and discretization}}
\\
\midrule

Geometry
&
$h$
&
Total coating thickness
&
mm
&
\revb{$2.0$}
&
Design input
\\

Discretization
&
$M$
&
No.\ of through-thickness \revb{homogenized subdomains}
&
--
&
\revb{$40$}
&
This study \revb{(adopted; verified Sect.~\ref{subsec:disc})}
\\

Time stepping
&
$\Delta t$
&
Time step
&
s
&
\revb{$0.25$}
&
This study \revb{(adopted; verified Sect.~\ref{subsec:disc})}
\\

Time horizon
&
$t_\mathrm{final}$
&
Final simulation time
&
s
&
\revb{$60$}
&
This study
\\

Coordinate
&
$z$
&
Through-thickness coordinate
&
\revb{mm}
&
\revb{$[0,2]$}
&
$z\in[0,h]$ \revb{; $0$ interface, $h$ free surface}
\\

\revb{Depth}
&
\revb{$\zeta$}
&
\revb{Normalized depth below free surface}
&
\revb{--}
&
\revb{$[0,1]$}
&
\revb{$\zeta=(h-z)/h$; used for all results}
\\

\midrule
\multicolumn{6}{l}{
\textbf{Thermo-viscoelasticity (Prony / generalized Maxwell)}
}
\\
\midrule

Viscoelasticity
&
$E_{\infty}(T)$
&
Long-term equilibrium modulus
&
\revb{--}
&
\revb{$0.60\,E_i(z,T)$}
&
Literature / calibration \revb{(benchmark)}
\\

Viscoelasticity
&
$E_m(T)$
&
$m$-th Prony modulus coefficient
&
\revb{--}
&
\revb{$E_{1,2,3}=0.20,\,0.12,\,0.08\,E_i$}
&
Literature / calibration \revb{(benchmark)}
\\

Viscoelasticity
&
$\tau_m$
&
$m$-th relaxation time
\revb{(constant in $T$; no shift factor)}
&
s
&
\revb{$\tau_{1,2,3}=5,\,50,\,500$}
&
Literature / calibration \revb{; ident.\ required}
\\

Viscoelasticity
&
$M_p$
&
Number of Prony terms
&
--
&
\revb{$3$}
&
Model choice
\\

Temperature
&
$T$
&
Prescribed temperature field
&
$^\circ$C
&
\revb{$20\!\rightarrow\!400\!\rightarrow\!20$}
&
Operating condition
\revb{(benchmark cycle, Fig.~\ref{fig:cycle})}
\\

\revb{Reference}
&
\revb{$T_\mathrm{ref}$}
&
\revb{Stress-free reference temperature}
&
\revb{$^\circ$C}
&
\revb{$20$}
&
\revb{Benchmark}
\\

\midrule
\multicolumn{6}{l}{
\revb{\textbf{Thermal expansion
(eigenstrain and coating--substrate mismatch)}}
}
\\
\midrule

\revb{Expansion}
&
\revb{$\alpha_c$}
&
\revb{Coating expansion coefficient
(uniform through the thickness in the benchmark)}
&
\revb{$10^{-6}$/K}
&
\revb{$16$}
&
\revb{Benchmark; ident.\ required}
\\

\revb{Expansion}
&
\revb{$\alpha_s$}
&
\revb{Substrate expansion coefficient}
&
\revb{$10^{-6}$/K}
&
\revb{$13$}
&
\revb{Benchmark; ident.\ required}
\\

\revb{Mismatch}
&
\revb{$\Delta\alpha$}
&
\revb{$\alpha_c-\alpha_s$, drives Eq.~\eqref{eq:mismatch}}
&
\revb{$10^{-6}$/K}
&
\revb{$3$}
&
\revb{Derived}
\\

\revb{Mismatch}
&
\revb{$\Delta T_{\max}$}
&
\revb{$T_{\max}-T_\mathrm{ref}$ over the benchmark cycle}
&
\revb{K}
&
\revb{$380$}
&
\revb{Derived (Fig.~\ref{fig:cycle}a)}
\\

\revb{Mismatch}
&
\revb{$\varepsilon^{\mathrm{mis}}_{\max}$}
&
\revb{$\Delta\alpha\,\Delta T_{\max}$, peak mismatch strain}
&
\revb{$10^{-3}$}
&
\revb{$1.14$}
&
\revb{Derived; plotted in Fig.~\ref{fig:cycle}b}
\\

\midrule
\multicolumn{6}{l}{
\textbf{Functionally graded material and modulation}
\revb{\textbf{(AlSi--hBN reference family)}}
}
\\
\midrule

FGM
&
$E_\mathrm{met}(T)$
&
Metallic-phase modulus
&
\revb{GPa}
&
\revb{$48$ ($20\,^\circ$C); $31$ ($400\,^\circ$C)}
&
Literature / material data
\revb{; porosity-inclusive~\cite{faraoun2006abradable}}
\\

FGM
&
$E_\mathrm{cer}(T)$
&
Ceramic-phase modulus
&
\revb{GPa}
&
\revb{$12$ ($20\,^\circ$C); $8$ ($400\,^\circ$C)}
&
Literature / material data
\revb{; porosity-inclusive~\cite{faraoun2006abradable}}
\\

FGM
&
$n_g$
&
Gradient exponent
&
--
&
\revb{$2.0$}
&
Design choice / tuning
\\

Modulation
&
\revb{$\Delta E_g$}
&
Modulation amplitude
&
\revb{GPa}
&
\revb{$2.0$ ($\approx4$--$17\%$ of local $E$)}
&
Deposition heterogeneity
\revb{\cite{lokachari2024columnar}; nanoindentation}
\\

Modulation
&
$\lambda_g$
&
Modulation wavelength
&
\revb{mm}
&
\revb{$0.4$}
&
Microstructure assumption
\revb{; pass-group (Sect.~\ref{subsec:fgm_mod})}
\\

\revb{Porosity}
&
\revb{$\phi$}
&
\revb{Total porosity (homogenized)}
&
\revb{\%}
&
\revb{$30$--$45$}
&
\revb{Embedded in effective moduli~\cite{liu2012thermoelastic}}
\\

\midrule
\multicolumn{6}{l}{\textbf{Damage model}}
\\
\midrule

Damage
&
$D$
&
Scalar damage variable ($0\le D\le1$)
&
--
&
\revb{$[0,1]$}
&
Model state
\\

Damage
&
$\sigma_\mathrm{crit}$
&
Critical stress threshold
&
\revb{MPa}
&
\revb{$80$}
&
Literature / calibration
\revb{; effective value (Sect.~\ref{subsec:damage})}
\\

Damage
&
$A$
&
Damage rate coefficient
&
\revb{s$^{-1}$}
&
\revb{$0.12$}
&
Literature / calibration
\revb{; anchored on~\cite{bertuol2025alsi}, $300\,^\circ$C}
\\

Damage
&
$m_d$
&
Damage exponent (overstress)
&
--
&
\revb{$2.1$}
&
Literature / calibration
\revb{; anchored on~\cite{bertuol2025alsi}}
\\

Damage
&
$n_d$
&
Damage exponent (saturation)
&
--
&
\revb{$3.2$}
&
Literature / calibration
\revb{; anchored on~\cite{bertuol2025alsi}}
\\

\revb{Screening metric}
&
$D_\mathrm{crit}$
&
\revb{Damage classification level}
&
--
&
\revb{$0.10$}
&
This study
\revb{; declared level, not a rupture criterion
(Sect.~\ref{subsec:variability})}
\\

\midrule
\multicolumn{6}{l}{
\revb{\textbf{Prescribed mechanical excitation}}
}
\\
\midrule

\revb{Excitation}
&
\revb{$\varepsilon^{\mathrm{app}}_0$}
&
\revb{Applied normal-strain amplitude (compressive)}
&
\revb{$10^{-3}$}
&
\revb{$2.11$}
&
\revb{Benchmark input, \emph{not} a contact solution}
\\

\revb{Excitation}
&
\revb{$n_p$}
&
\revb{Number of approach pulses per cycle}
&
\revb{--}
&
\revb{$2$}
&
\revb{Benchmark waveform (Sect.~\ref{subsec:loading})}
\\

\revb{Excitation}
&
\revb{$t_r$}
&
\revb{Pulse ramp time}
&
\revb{s}
&
\revb{$2$}
&
\revb{Benchmark waveform (Sect.~\ref{subsec:loading})}
\\

\revb{Excitation}
&
\revb{$t_d$}
&
\revb{Pulse hold (plateau) time}
&
\revb{s}
&
\revb{$4$}
&
\revb{Benchmark waveform (Sect.~\ref{subsec:loading})}
\\

\midrule
\multicolumn{6}{l}{
\textbf{Probabilistic tolerance analysis (Monte Carlo)}
}
\\
\midrule

Tolerances
&
$\delta\mathbf{u}$
&
Vector of geometric deviations
&
mm
&
\revb{---}
&
Manufacturing / assembly specifications
\\

Uncertainty
&
$\delta u_j\sim\mathcal{N}(0,\sigma_{u_j}^2)$
&
Example tolerance distribution
&
\revb{mm}
&
\revb{$\sigma_{u_j}=0.02$}
&
Modeling assumption
\revb{; $\pm3\sigma=\pm0.06$~mm band}
\\

\revb{Tolerance map}
&
\revb{$L_{\mathrm{eff}}$}
&
\revb{Effective compliance length of Eq.~\eqref{eq:tolmap}}
&
\revb{mm}
&
\revb{$75$}
&
\revb{Reduced-order model input; benchmark}
\\

Monte Carlo
&
$N$
&
Number of realizations
&
--
&
\revb{$800$}
&
This study
\revb{(nested convergence, Sect.~\ref{subsec:prob})}
\\

\revb{Surrogate}
&
\revb{$p$}
&
\revb{Polynomial response-surface order}
&
\revb{--}
&
\revb{$3$}
&
\revb{$400+400$ runs (Fig.~\ref{fig:mcpce})}
\\

Outputs
&
$\sigma_{\max}$
&
Maximum stress statistic
&
\revb{MPa}
&
\revb{---}
&
Post-processing metric
\\

\revb{Outputs}
&
\revb{$P_e$}
&
\revb{Probability of exceeding $D_\mathrm{crit}$}
&
\revb{--}
&
\revb{---}
&
\revb{Post-processing metric}
\\

\bottomrule
\end{tabular}

\endgroup
\end{table}

\subsubsection{Physical system versus solved numerical domain}
\label{subsec:domain}
Fig.~\ref{fig:config} distinguishes two levels that must be kept separate.
Panel~(a) is the physical system: sealing teeth
carried on one member face the coated counter-face across a nominal running
clearance, and relative radial motion may close that clearance and create a
localized rubbing region. Normal pressure, tangential traction and frictional
heat are genuine physical quantities at that level. Panel~(b) is the
solved domain: a local through-thickness coating column of total
thickness $h$, divided into $M$ perfectly bonded homogenized subdomains, resting
on a substrate that enters the formulation only as a kinematic
thermal-expansion constraint through the mismatch of Eq.~\eqref{eq:mismatch}; no
substrate modulus, thickness or force balance is solved.

The subdomain boundaries are continuity conditions, not sliding interfaces:
displacement and traction continuity is enforced through the thickness, so the
concern that friction between internal layers would violate continuity does not
arise, because no such interface exists. The discretization is therefore defined
by the number $M$ of homogenized subdomains and by the time step $\Delta t$, not
by a two- or three-dimensional finite-element contact mesh; no contact elements,
gap-closure algorithm or friction law are present in the solver.
Table~\ref{tab:scope} lists, quantity by quantity, what is physically present,
what is prescribed as an input, what is computed, and what is excluded.

\begin{table}[!t]
\centering
\caption{\revb{Model scope. Excluded quantities appear nowhere in the results,
figures or conclusions of this paper.}}
\label{tab:scope}
\footnotesize
\renewcommand{\arraystretch}{1.15}
\setlength{\tabcolsep}{5pt}
\begin{tabular}{@{}p{\dimexpr 0.3433\linewidth-2\tabcolsep\relax}p{\dimexpr 0.1716\linewidth-2\tabcolsep\relax}p{\dimexpr 0.1567\linewidth-2\tabcolsep\relax}p{\dimexpr 0.3284\linewidth-2\tabcolsep\relax}@{}}
\toprule
\textbf{Quantity / phenomenon} & \textbf{Present in physical system} &
\textbf{Solved here?} & \textbf{Representation in the solver} \\
\midrule
Normal approach / interference & Yes & No &
Converted to prescribed $\varepsilon^{\mathrm{app}}(t)$ (input) \\
Contact pressure $p_n(z,t)$ & Yes & No & Not computed; enters via the input history \\
Tangential traction $q_t(z,t)$ & Yes & No & Not resolved; excluded \\
Coulomb friction coefficient $\mu$ & Yes & No & \textbf{Absent from the formulation} \\
Frictional heat generation & Yes & No & Effect only via prescribed $T(z,t)$ \\
Material removal / wear depth & Yes & No & \textbf{Not solved; not reported} \\
Rotor dynamics, leakage flow & Yes & No & Outside the reduced domain \\
Individually resolved pores, splats, cracks & Yes & No &
Homogenized effective properties only \\
\midrule
Temperature history $T(z,t)$ & --- & Input & Prescribed benchmark \\
Normal-strain history $\varepsilon^{\mathrm{app}}(t)$ & --- & Input & Prescribed benchmark \\
Thermal eigenstrain, expansion mismatch & --- & Solved & Eqs.~\eqref{eq:strainsplit}--\eqref{eq:mismatch} \\
Viscoelastic stress $\sigma(z,t)$ & --- & Solved & Computed output \\
Scalar damage $D(z,t)$ & --- & Solved & Computed output \\
Exceedance probability $P_e$ & --- & Solved & Monte Carlo output \\
\bottomrule
\end{tabular}
\end{table}

\subsubsection{Prescribed loading histories}
\label{subsec:loading}
Two prescribed histories drive the column, both illustrated in
Fig.~\ref{fig:cycle}:

The thermal history is a trapezoidal benchmark cycle: heating from
$T_0=20\,^\circ$C to $T_{\max}=400\,^\circ$C over $0\le t\le20$~s, an isothermal
dwell at $400\,^\circ$C for $20\le t\le40$~s, and cooling back to
$20\,^\circ$C for $40\le t\le60$~s. The stress-free reference temperature is
$T_{\mathrm{ref}}=20\,^\circ$C. Within each subdomain the temperature is uniform
at each instant. This cycle is a reproducible numerical benchmark chosen to
exercise the full range of viscoelastic relaxation and thermal mismatch; it is
not an engine duty cycle, and no claim of duty-cycle representativeness
is made.

The mechanical history is an equivalent normal-strain history
$\varepsilon^{\mathrm{app}}(t)$ applied at the coating boundary, representing the
normal approach of the sealing member in a cyclic manner synchronized with the
thermal dwell. It is an input, not a contact solution: the conversion from
physical interference to equivalent prescribed strain is an assumption of the
reduced-order model, and the amplitude
$\varepsilon^{\mathrm{app}}_0$ is a declared benchmark quantity
(Table~\ref{tab:param_summary}). \revb{Its basis is stated explicitly rather than
left implicit. The value $2.11\times10^{-3}$ was selected so that the peak
equivalent stress generated by the benchmark falls within the stress range over
which the damage parameters of Section~\ref{subsec:calib} are identified; below
roughly $1.6\times10^{-3}$ the column never reaches $\sigma_\mathrm{crit}$ and no
damage is produced at all, which would leave the damage model unexercised. The
resulting stress level is therefore a designed property of the benchmark and is
not presented as an independent prediction.}

\revb{The waveform is stated here in full so that it can be reproduced without
access to the driver script. It consists of $n_p=2$ identical trapezoidal
compressive pulses placed inside the isothermal dwell, each built from a linear
ramp of duration $t_r=2$~s, a plateau of duration $t_d=4$~s at the amplitude
$-\varepsilon^{\mathrm{app}}_0$, and a linear return ramp of duration $t_r$:}
\begin{equation}
  \revb{\varepsilon^{\mathrm{app}}(t)=-\,\varepsilon^{\mathrm{app}}_{0}
  \sum_{k=1}^{n_p}\Lambda\!\left(\frac{t-t_k}{t_r},\frac{t_d}{t_r}\right),
  \qquad
  \varepsilon^{\mathrm{app}}_{0}=2.11\times10^{-3},
  \qquad
  t_1=18~\mathrm{s},\; t_2=30~\mathrm{s},}
  \label{eq:appstrain}
\end{equation}
\revb{where $\Lambda$ denotes the unit trapezoid that rises linearly from $0$ to
$1$ on $[0,1]$, remains at $1$ on $[1,1+t_d/t_r]$ and returns linearly to $0$ on
$[1+t_d/t_r,2+t_d/t_r]$. The strain is therefore zero for $t<18$~s and for
$t>38$~s, and the two plateaux occupy $20\le t\le24$~s and $32\le t\le36$~s. The
sign convention is that compression is negative, consistent with a normal
approach of the sealing member.}

\revb{A third strain contribution is generated internally rather than prescribed,
and is therefore not one of the two input histories above. Because
the coating is bonded to a substrate of different thermal expansion coefficient,
the constraint produces an expansion-mismatch strain governed by the difference
$\Delta\alpha=\alpha_c-\alpha_s$, formally defined in Eq.~\eqref{eq:mismatch}
below. Anticipating that definition, and noting that $\alpha_c$ is uniform
through the thickness in the benchmark so that
$\varepsilon^{\mathrm{mis}}$ is a function of time only, its peak value over the
cycle is}
\begin{equation}
  \revb{\varepsilon^{\mathrm{mis}}_{\max}
  =\Delta\alpha\,\Delta T_{\max}
  =(16-13)\times10^{-6}\,\mathrm{K^{-1}}\times(400-20)\,\mathrm{K}
  =1.14\times10^{-3},}
  \label{eq:mismax}
\end{equation}
\revb{which is the plateau of the dashed curve in Fig.~\ref{fig:cycle}b. The
applied-strain amplitude and the peak mismatch strain are therefore of the same
order, $\varepsilon^{\mathrm{app}}_0/\varepsilon^{\mathrm{mis}}_{\max}=1.85$, so
neither contribution dominates the mechanical strain of
Eq.~\eqref{eq:mechstrain}. Both histories are plotted in
Fig.~\ref{fig:cycle}b.}

The through-thickness coordinate is measured from the coating--substrate
interface, $z=0$, to the free surface, $z=h$, so that the metallic-rich end of
the gradient lies at $z=0$ and the compliant, ceramic/dislocator-rich end lies
at the free surface, consistently with Eq.~\eqref{eq:fgm_modulus}. Results are
reported against the normalized depth below the free surface,
$\zeta=(h-z)/h\in[0,1]$, so that $\zeta=0$ is the free surface and $\zeta=1$ the
interface. Subdomains are indexed $i=1,\dots,M$ from the free surface inwards.
}

\begin{figure}[!t]
\centering
\includegraphics[width=\textwidth,height=0.78\textheight,keepaspectratio]{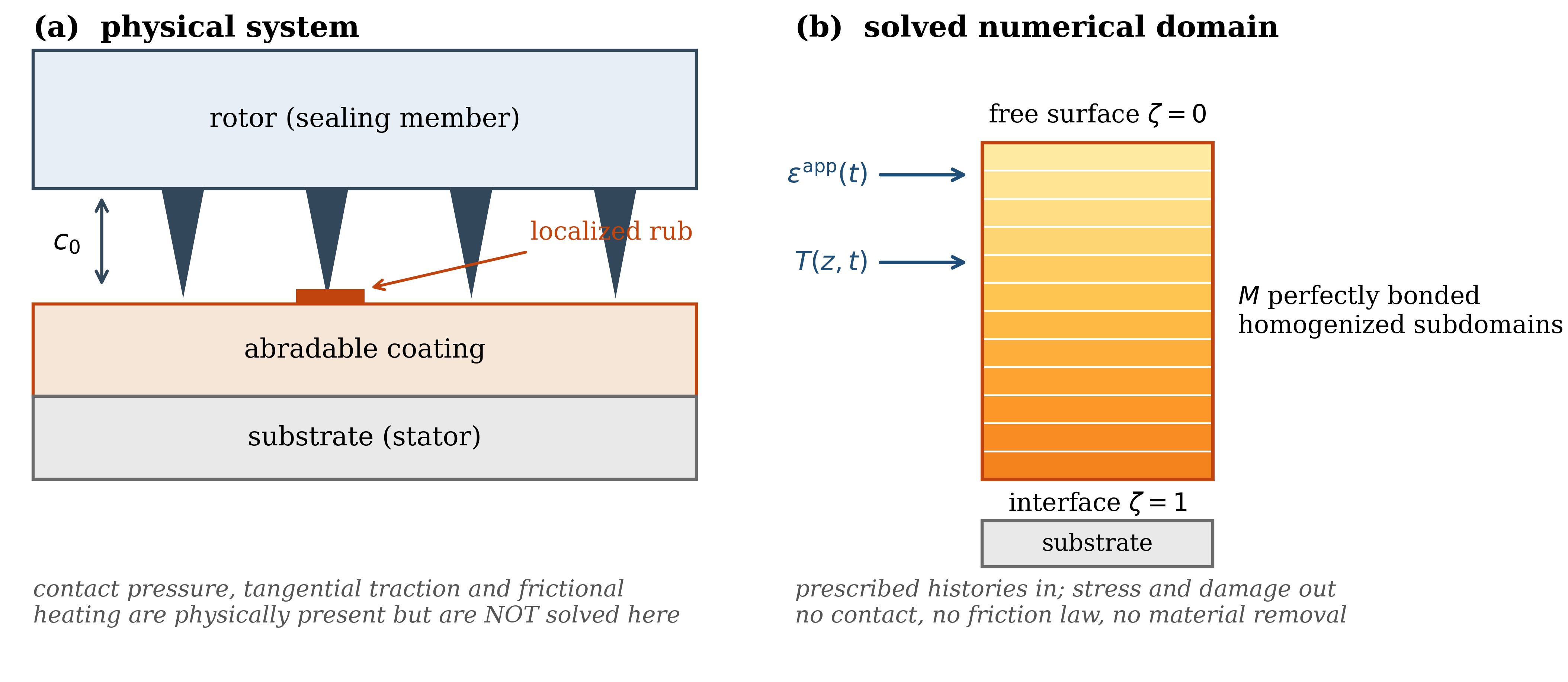}
\caption{\revb{Separation of the physical system from the solved numerical
domain. (a)~In the physical labyrinth-seal stage, relative radial approach may
close the clearance $c_0$ and create a localized rub region in which contact
pressure, tangential traction and frictional heating are real phenomena.
(b)~The domain actually solved is a local through-thickness coating column of $M$
perfectly bonded homogenized subdomains on a substrate that enters only through
the thermal expansion mismatch, driven by a prescribed equivalent normal-strain
history $\varepsilon^{\mathrm{app}}(t)$ and a prescribed temperature history
$T(z,t)$.}}
\label{fig:config}
\end{figure}
\begin{figure}[!t]
\centering
\includegraphics[width=\textwidth,keepaspectratio]{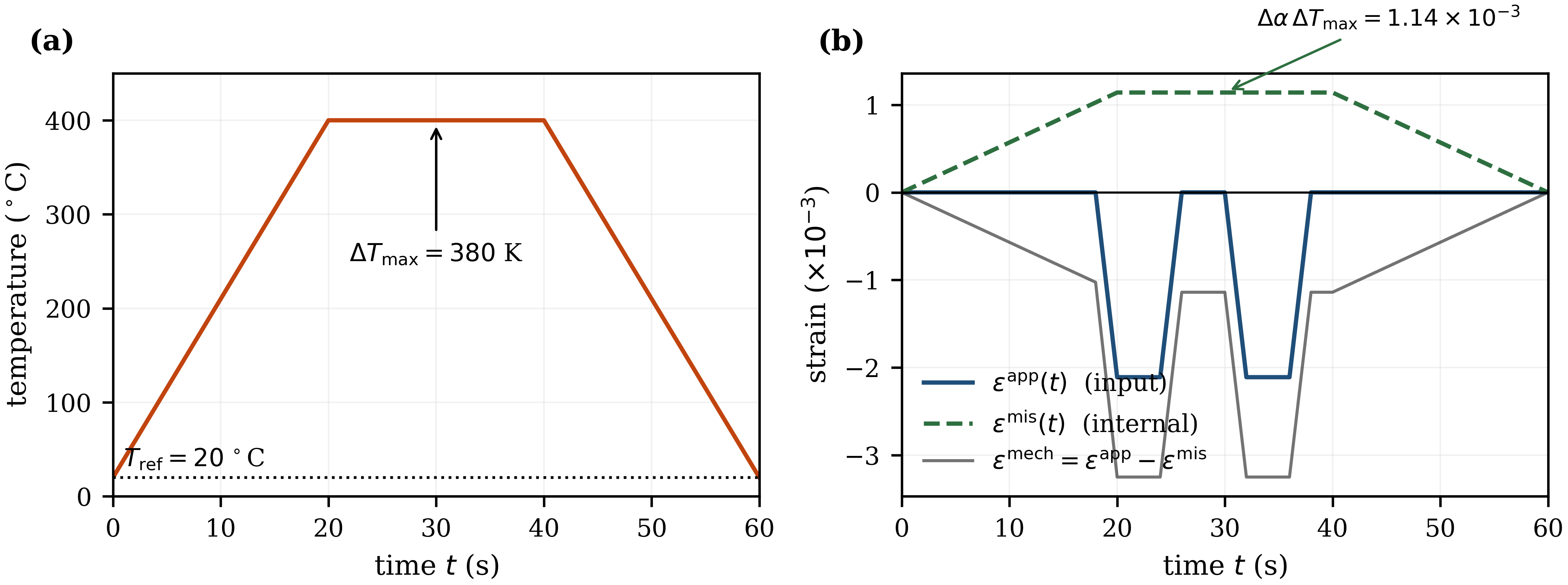}
\caption{\revb{Prescribed loading benchmark. (a)~The trapezoidal thermal cycle
runs $20\rightarrow400\rightarrow20\,^\circ$C over $60$~s with a $20$-s dwell,
and the stress-free reference temperature is
$T_{\mathrm{ref}}=20\,^\circ$C. (b)~The equivalent normal-strain history
$\varepsilon^{\mathrm{app}}(t)$ is an input representing cyclic normal approach,
whereas the coating--substrate expansion mismatch strain
$\varepsilon^{\mathrm{mis}}$ is generated internally by the thermal eigenstrain
of Eqs.~\eqref{eq:strainsplit}--\eqref{eq:mismatch}. The applied waveform is the
two-pulse trapezoid of Eq.~\eqref{eq:appstrain} with amplitude
$\varepsilon^{\mathrm{app}}_0=2.11\times10^{-3}$, ramp time $t_r=2$~s and plateau
$t_d=4$~s; the mismatch plateau is
$\Delta\alpha\,\Delta T_{\max}=3\times10^{-6}\,\mathrm{K^{-1}}\times380\,\mathrm{K}
=1.14\times10^{-3}$, i.e.\ exactly the value implied by the coefficients
$\alpha_c=16\times10^{-6}$~K$^{-1}$ and $\alpha_s=13\times10^{-6}$~K$^{-1}$ of
Table~\ref{tab:param_summary}. The cycle is a reproducible verification
benchmark, not an engine duty cycle.}}
\label{fig:cycle}
\end{figure}
\subsection{Modeling assumptions}
\label{subsec:assumptions}

The abradable coating is modeled as a layered, functionally graded medium
subjected to thermo-mechanical loading representative of turbine labyrinth seal
operation. Within the temperature range considered
\revb{($20$--$400\,^\circ$C, Section~\ref{subsec:loading})}, the material
response is assumed to be predominantly thermo-viscoelastic. Progressive
material degradation is described using a scalar isotropic damage variable,
while irreversible viscoplastic effects are neglected.

\revb{Thermal expansion is retained explicitly. The total
strain is decomposed into a mechanical part and a thermal eigenstrain, and the
coating--substrate expansion mismatch is included explicitly
(Section~\ref{subsec:viscoelastic}), so that thermal stress is generated by the
model rather than prescribed. In the benchmark parameter set the coating
expansion coefficient is taken as uniform through the thickness,
$\alpha_c(z,T)\equiv\alpha_c=16\times10^{-6}$~K$^{-1}$, so that the mismatch
strain of Eq.~\eqref{eq:mismatch} is a function of time alone; the general
formulation retains the depth and temperature dependence
$\alpha_c(z,T)$ and admits a graded coefficient without any change to the
solution algorithm. The expansion coefficients are declared benchmark quantities
requiring material identification (Table~\ref{tab:param_summary}).}

\begin{figure}[!t]
\centering
\includegraphics[width=\textwidth,height=0.78\textheight,keepaspectratio]{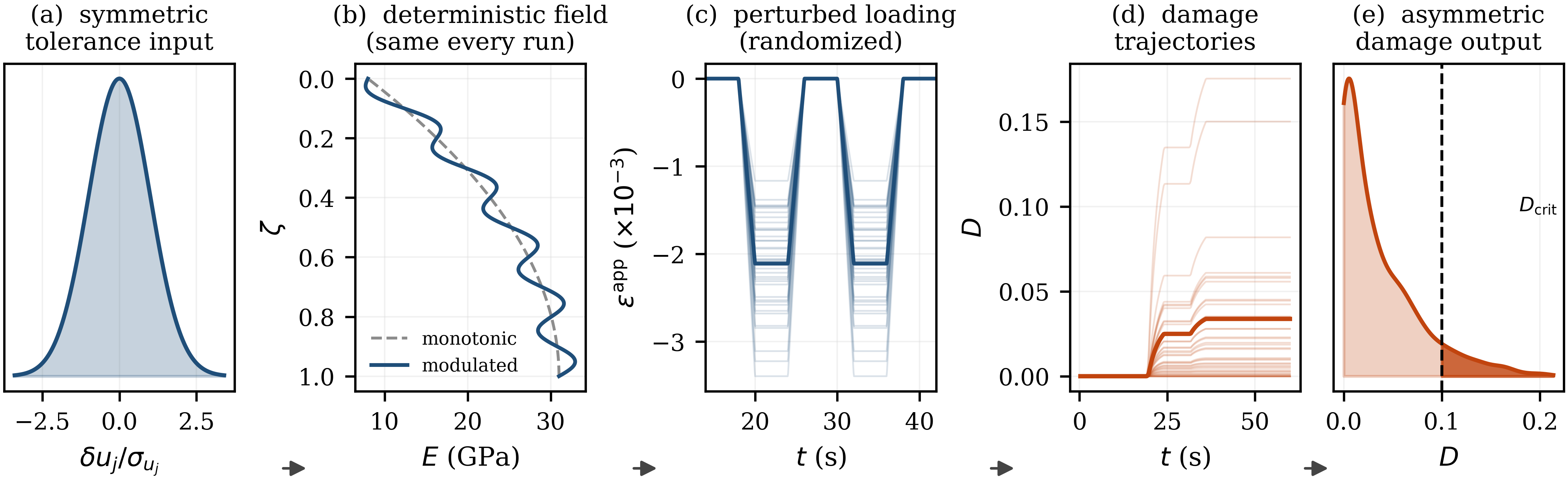}
\caption{\revb{Propagation of variability along the computational pipeline.
(a)~The sampled geometric tolerance is a symmetric Gaussian,
$\delta u_j\sim\mathcal{N}(0,\sigma_u^2)$ with $\sigma_u=0.02$~mm and a
$\pm3\sigma_u$ band. (b)~The deterministic monotonic and periodically modulated
property fields of Eq.~\eqref{eq:fgm_modulus} (wavelength $\lambda_g$); these are
identical in every realization and are not randomized in the present uncertainty
model. (c)~The ensemble of perturbed equivalent loading histories obtained from
the tolerance input through Eq.~\eqref{eq:tolmap}, which is the only randomized
quantity. (d)~The coupled solver returns a family of damage trajectories, the
modulated mean lying systematically above the monotonic mean and reaching
$0.03376$ at $t=60$~s against $0.02135$ for the monotonic case. (e)~The
end-of-cycle damage distribution is asymmetric, with the shaded
tail beyond the declared classification level $D_\mathrm{crit}=0.10$ giving the
exceedance probability $P_e=0.076$, against $0.035$ for the
monotonic gradient at the same level. A symmetric geometric input therefore maps into an asymmetric
damage output. Because the mapping is not measure-preserving, the ensemble mean
$0.034$ is not itself the nominal-geometry response; what a deterministic analysis
at nominal geometry cannot deliver is the exceedance mass beyond
$D_\mathrm{crit}$, which is a property of the distribution and not of any single
realization.}}
\label{fig:var_propagation}
\end{figure}

The functional gradient varies continuously through the coating thickness and
incorporates a periodic modulation to represent deposition-induced
microstructural heterogeneity. Thermal loading is prescribed and assumed uniform
within each \revb{homogenized subdomain}. Manufacturing and assembly tolerances
are treated separately within a probabilistic framework.
\revb{Fig.~\ref{fig:var_propagation} shows how the statistical content of the
problem is transformed along the computational pipeline. A symmetric Gaussian
tolerance input (panel~a) perturbs the amplitude of the prescribed equivalent
strain through Eq.~\eqref{eq:tolmap}, giving the ensemble of loading histories of
panel~c, while the deterministic through-thickness property field of panel~b is
identical in every realization. The perturbed histories are propagated through
the coupled solver to give a family of damage trajectories (panel~d) and
post-processed into the end-of-cycle damage distribution (panel~e). The transformation is not
measure-preserving: the symmetric geometric input produces an asymmetric
damage output whose exceedance probability beyond
$D_\mathrm{crit}$ is precisely the quantity that a deterministic analysis at
nominal geometry cannot reproduce.}

\subsection{Model parameter summary and uncertainty definition}
\label{sec:param_summary}

Table~\ref{tab:param_summary} summarizes the mechanical, thermal, numerical, and
probabilistic inputs used in the deterministic and Monte Carlo analyses,
including units\revb{, numerical values, origin, and declared status}. The
selection of parameters is consistent with the literature on functionally graded
materials, continuum damage mechanics, and structural reliability
analysis~\cite{suresh1998fundamentals,lemaitre2005engineering,%
ditlevsen1996reliability,yildirim2011periodic}.

\subsection{Thermo-viscoelastic formulation \revb{with thermal eigenstrain}}
\label{subsec:viscoelastic}

The thermo-viscoelastic behavior of the coating is described using a linear
formulation based on a Prony series representation of the relaxation modulus
\revb{\cite{schapery1969nonlinear,simo1996schapery,ferry1980viscoelastic}}.
The temperature-dependent relaxation modulus is expressed as
\revb{\cite{ferry1980viscoelastic,brinson2008polymer}}
\begin{equation}
  E(t,T) = E_{\infty}(T)
          + \sum_{m=1}^{M_p} E_{m}(T)
            \exp\!\left(-\frac{t}{\tau_{m}}\right),
  \label{eq:prony}
\end{equation}
where $E_{\infty}(T)$ denotes the long-term elastic modulus, $E_{m}(T)$ are the
Prony coefficients, and $\tau_{m}$ are the relaxation times.
\revb{Equation~\eqref{eq:prony} is the classical generalized Maxwell (Prony)
representation of linear viscoelasticity and is not derived here. Its
non-isothermal implementation is stated explicitly, because the benchmark spans
$20$--$400\,^\circ$C and the constitutive response cannot otherwise be
reproduced. The temperature enters through the instantaneous modulus only: the
phase moduli $E_\mathrm{met}$ and $E_\mathrm{cer}$ of Table~\ref{tab:param_summary}
are interpolated linearly in $T$ between their $20$ and $400\,^\circ$C values,
$E_i(z,T)$ follows from Eq.~\eqref{eq:fgm_modulus}, and the Prony weights are
constant fractions of $E_i(z,T)$, namely $E_\infty=0.60\,E_i$ and
$E_{1,2,3}=0.20,0.12,0.08\,E_i$. The relaxation times are held constant at
$\tau_{1,2,3}=5,50,500$~s; no time--temperature shift factor of WLF or Arrhenius
type is applied, so the model is not thermorheologically simple and no master
curve is constructed. The hereditary integral is advanced by the standard
recursive update of the internal variables, $q_m^{\,n+1}=e^{-\Delta t/\tau_m}q_m^{\,n}
+E_m(T^{n+1})\,\tau_m\bigl(1-e^{-\Delta t/\tau_m}\bigr)\Delta\varepsilon^{\mathrm{mech}}/\Delta t$,
with the moduli evaluated at the temperature of the current step, which is exact
for a strain varying linearly within the step at constant temperature and
first-order accurate in the temperature variation across it. Introducing a shift
factor identified from dynamic mechanical analysis is a defined refinement, listed
in Section~\ref{sec:implications_limits_prospects}.}

\revb{Because the coating is heated while bonded to a substrate, the total strain
is decomposed into a mechanical (stress-producing) part and a thermal
eigenstrain,}
\begin{equation}
  \revb{\varepsilon^{\mathrm{tot}}(z,t)
   = \varepsilon^{\mathrm{mech}}(z,t) + \varepsilon^{\mathrm{th}}(z,t),
   \qquad
   \varepsilon^{\mathrm{th}}(z,t) = \alpha_c(z,T)\,
   \bigl[T(z,t)-T_\mathrm{ref}\bigr],}
  \label{eq:strainsplit}
\end{equation}
\revb{and the constraint exerted by the substrate introduces an expansion
mismatch strain}
\begin{equation}
  \revb{\varepsilon^{\mathrm{mis}}(z,t)
   = \bigl[\alpha_c(z,T)-\alpha_s(T)\bigr]\,
     \bigl[T(z,t)-T_\mathrm{ref}\bigr],}
  \label{eq:mismatch}
\end{equation}
\revb{so that the mechanical strain driving the constitutive response is}
\begin{equation}
  \revb{\varepsilon^{\mathrm{mech}}(z,t)
   = \varepsilon^{\mathrm{app}}(t) - \varepsilon^{\mathrm{mis}}(z,t).}
  \label{eq:mechstrain}
\end{equation}
\revb{Equations~\eqref{eq:strainsplit}--\eqref{eq:mechstrain} are the standard
thermo-elastic strain decomposition applied to a constrained coating
layer~\cite{brinson2008polymer,liu2012thermoelastic}. One consequence of the
reduced-order form should be stated plainly, because it bounds how the results
may be read: the prescribed normal approach and the coating--substrate expansion
mismatch are combined into a single equivalent stress-producing strain. In a full
continuum treatment these are tensorially distinct, the former being a
through-thickness compression and the latter an in-plane biaxial mismatch stress.
The present formulation does not resolve that distinction, and it is therefore an
equivalent one-dimensional surrogate for the constrained response rather than a
complete thermoelastic constraint model. With
$\alpha_c=\alpha_s$ and $\varepsilon^{\mathrm{app}}=0$ the formulation reduces to
stress-free expansion, which is used as a closed-form unit test in
Section~\ref{subsec:disc}.}

The viscoelastic stress response is obtained through the Boltzmann superposition
(hereditary) integral of the relaxation modulus with the
\revb{mechanical} strain-rate history
\revb{\cite{schapery1969nonlinear,wineman2000mechanical}},
\begin{equation}
  \sigma(z,t) = \int_{0}^{t} E\bigl(t-\xi,T\bigr)\,
  \frac{d\varepsilon^{\mathrm{mech}}(z,\xi)}{d\xi}\,d\xi .
  \label{eq:convolution}
\end{equation}
This formulation captures the memory effects and delayed stress response under
transient thermo-mechanical loading and provides the basis for coupling with
damage evolution.

\subsection{Progressive damage model}
\label{subsec:damage}

Material degradation is represented by a scalar damage variable
$D(t)\in[0,1]$, accounting for stiffness reduction due to micro-cracking and
matrix degradation \revb{\cite{kachanov1986damage,lemaitre2005engineering}}.
Although damage may be anisotropic at the microscale, a scalar formulation is
adopted to ensure numerical robustness and computational efficiency at the
coating scale.

\revb{The validity of this coating-scale scalar description rests on a
separation-of-scales (representative elementary volume, REV) argument, which is
made explicit here for the heterogeneous graded system
considered~\cite{kanit2003rve}. Three characteristic lengths must be ordered:
the microstructural feature size $\ell_\mu$ (splats, pores, dislocator
particles; $\ell_\mu\approx5$--$30~\mu$m for as-sprayed
material~\cite{faraoun2006abradable}), the subdomain thickness $h/M=50~\mu$m
used in the adopted discretization, and the modulation wavelength
$\lambda_g=0.4$~mm. The adopted configuration satisfies
$\ell_\mu\lesssim h/M<\lambda_g\ll h$, while the modulation wavelength is
resolved by eight subdomains per period so that the macroscopic property
oscillation is not aliased by the homogenization. The margin in the first
inequality should be stated honestly rather than overclaimed: $h/M=50~\mu$m
exceeds the coarse end of the feature range ($30~\mu$m) by a factor of only
$1.7$, and the fine end ($5~\mu$m) by a factor of $10$. A subdomain is therefore
not a deterministic REV in the classical sense of containing an
arbitrarily large feature population. The two roles must therefore be kept apart. The
effective properties assigned to each subdomain are {assumed} to have been
homogenized over an underlying representative volume of the sprayed
microstructure, in the statistical sense of Kanit et al.~\cite{kanit2003rve};
$h/M$ is a numerical discretization length whose function is to resolve the
macroscopic modulation, and it is not itself evidence that such a volume has been
sampled. The $N$ realizations of Section~\ref{subsec:prob} sample geometric
tolerances only and quantify no microstructural apparent-property variance.
Establishing REV convergence for a specific coating, by the ensemble criterion
of~\cite{kanit2003rve} applied to measured microstructures, is a separate
material-identification task and is listed as such in
Section~\ref{sec:implications_limits_prospects}. The scalar variable is therefore defined at the
REV/subdomain scale, where the randomly oriented splat and pore population
justifies quasi-isotropy of the degradation, and it is used only for macroscopic
stiffness degradation; it is not claimed to represent sub-REV anisotropic crack
populations and is not used to predict crack paths. Should the subdomain
thickness approach the microstructural feature size, the REV assumption would
degrade and an anisotropic or nonlocal description would be required; this
boundary of validity is stated in
Section~\ref{sec:implications_limits_prospects}.}

Damage evolution is governed by a stress-driven kinetic law of the
Kachanov--Lemaitre type
\revb{\cite{kachanov1986damage,lemaitre2005engineering}},
\begin{equation}
  \frac{dD}{dt}
  = A\left\langle \frac{\revb{\sigma_{\mathrm{eq}}}(z,t)}{\sigma_\mathrm{crit}}-1 \right\rangle^{m_d}
    (1-D)^{n_d},
  \label{eq:damage_law}
\end{equation}
where $A$, $m_d$, and $n_d$ are material parameters, $\sigma_\mathrm{crit}$ is a
critical stress threshold, and $\langle\cdot\rangle$ denotes the Macaulay
bracket. \revb{The sign convention is fixed here and used without exception. Compression
is negative. With $\alpha_c>\alpha_s$ and heating, Eq.~\eqref{eq:mismatch} gives
$\varepsilon^{\mathrm{mis}}>0$, so Eq.~\eqref{eq:mechstrain} makes the mismatch
term add to the prescribed compressive approach rather than oppose it: both
contributions drive $\varepsilon^{\mathrm{mech}}$ negative, and the constitutive
stress of Eq.~\eqref{eq:convolution} is compressive throughout the loaded phase
of the benchmark. A small positive excursion of a few MPa occurs during
unloading, as viscoelastic recovery returns the column through zero, and it
remains far below $\sigma_\mathrm{crit}$ so that it never activates
Eq.~\eqref{eq:damage_law}. Two distinct measures are therefore used and are kept
distinct throughout: the signed constitutive stress $\sigma(z,t)$ of
Eq.~\eqref{eq:convolution}, and the non-negative equivalent stress}
\begin{equation}
  \revb{\sigma_{\mathrm{eq}}(z,t)\equiv|\sigma(z,t)| ,
   \qquad
   \sigma_{\max}(z)\equiv\max_t\,\sigma_{\mathrm{eq}}(z,t),}
  \label{eq:seq}
\end{equation}
\revb{which drives Eq.~\eqref{eq:damage_law} and carries every stress statistic
reported in Section~\ref{sec:results_discussion}. The same convention is
implemented in the solver and used in all figures.} Damage accumulates progressively when the local stress exceeds the
critical value and saturates as $D$ approaches unity, allowing the model to
capture nonlinear degradation under cyclic thermo-mechanical loading.
\revb{Equation~\eqref{eq:damage_law} is explicitly phenomenological: it
describes macroscopic stiffness loss and is not a mechanistic representation of
individual microstructural damage processes. One structural difference from the
classical creep-rupture form of the Kachanov--Lemaitre family should be stated
explicitly, because it changes the qualitative behaviour of the solution. The
classical form carries the damage factor with a {negative} exponent,
$(1-D)^{-k}$, which accelerates as $D$ grows and produces a finite rupture time.
Equation~\eqref{eq:damage_law} carries it with a {positive} exponent
$n_d>0$, so the rate decreases monotonically with accumulated damage and
$D\to1$ only asymptotically: the law is saturating rather than
rupture-generating. This is a deliberate choice for a stiffness-degradation
descriptor at the coating scale, where the quantity of interest is the loss of
load-carrying capacity within a bounded cycle rather than a rupture time.
It has two consequences that are used throughout: $D_\mathrm{crit}$ is a
declared classification level and not a rupture criterion, and the
identified exponents are not directly comparable with exponents tabulated for
the negative-exponent creep-rupture form.}

\revb{A single value of the critical threshold, $\sigma_\mathrm{crit}=80$~MPa, is
used through the coating thickness even though the modulus is graded and
modulated. This is a consequence of the effective-homogenized-continuum
assumption adopted in Section~\ref{subsec:material}: the modulated FGM
microstructure is not resolved at the scale of individual pores or phases, and
its influence is represented through the spatial variation of the {effective}
properties. The threshold therefore corresponds to the macroscopic onset of
damage in the homogenized medium rather than to the local initiation threshold of
each constituent. Two consequences follow. First, because the stress field, and
not the threshold, carries the spatial heterogeneity, the predicted hot-spot
locations are governed by the stress concentrations induced by modulation, which
is precisely the mechanism under investigation. Second, this approximation is
appropriate for the global structural response but not for resolving local
microstructural damage initiation: a spatially varying threshold
$\sigma_\mathrm{crit}(z,T,\phi)$, informed by the local porosity and temperature,
would be required for that purpose and is identified as a model extension in
Section~\ref{sec:implications_limits_prospects}.}

\subsection{Functionally graded material with periodic microstructural
modulation}
\label{subsec:fgm_mod}

The spatial variation of the elastic modulus through the coating thickness is
defined to reduce global stress discontinuities while accounting for local
heterogeneities induced by the deposition process. The effective modulus in
\revb{subdomain} $i$ at position $z$ is expressed as
\begin{equation}
  E_{i}(z,T)
  = E_\mathrm{met}(T)
  + \bigl[E_\mathrm{cer}(T)-E_\mathrm{met}(T)\bigr]\!
    \left(\frac{z}{h}\right)^{n_g}
  + \revb{\Delta E_{g}}\sin\!\left(\frac{2\pi z}{\lambda_{g}}\right),
  \label{eq:fgm_modulus}
\end{equation}
where $E_\mathrm{met}(T)$ and $E_\mathrm{cer}(T)$ are the temperature-dependent
effective moduli of the metallic-rich and dislocator-rich ends, respectively.
The coordinate $z$ denotes the through-thickness position, $h$ is the total
coating thickness, $n_g$ is the functional gradient exponent,
\revb{$\Delta E_{g}$} is the amplitude of the sinusoidal modulation, and
$\lambda_{g}$ is the characteristic wavelength of the microstructural
heterogeneity. \revb{The modulation amplitude is denoted $\Delta E_g$ (units of
GPa) and must not be confused with the strain measures
$\varepsilon^{\mathrm{app}}$, $\varepsilon^{\mathrm{th}}$,
$\varepsilon^{\mathrm{mis}}$ and $\varepsilon^{\mathrm{mech}}$ of
Section~\ref{subsec:viscoelastic}.}

This representation enables simultaneous modeling of the global functional
gradient and local stiffness fluctuations, which are responsible for stress
localization and premature damage initiation
\revb{\cite{yildirim2011periodic,chen2021modulation}}.

\revb{The sinusoidal term in Eq.~\eqref{eq:fgm_modulus} has a direct
manufacturing origin. Thermal-spray deposition builds the coating pass by pass:
each torch raster deposits a band of splats, and the periodic alternation of
freshly deposited, well-melted material with inter-pass boundaries, oxide
stringers, and locally elevated porosity produces a quasi-periodic
through-thickness oscillation of density and
stiffness~\cite{lokachari2024columnar,huang2024review}. The wavelength
$\lambda_g$ therefore corresponds to the thickness deposited per raster group
(here $\lambda_g=0.4$~mm, i.e., groups of passes of roughly $10$--$50~\mu$m
each), and the amplitude $\Delta E_g$ quantifies the resulting stiffness
contrast. Both parameters are experimentally measurable: $\lambda_g$ can be
extracted from image analysis of polished cross-sections through the periodicity
of porosity and oxide banding, while $\Delta E_g$ can be identified from
through-thickness nanoindentation modulus profiles or spatially resolved
ultrasonic measurements; the Prony parameters follow from stress-relaxation or
dynamic mechanical analysis at the temperatures of
interest~\cite{ferry1980viscoelastic}, and the expansion coefficients from
dilatometry. In the present study $\Delta E_g$ and $\lambda_g$ are adopted as
declared benchmark values (Table~\ref{tab:param_summary}); a dedicated
experimental identification campaign is part of the validation pathway of
Section~\ref{sec:implications_limits_prospects}. The idealization of the
modulation as a single harmonic retains the dominant Fourier component of the
measured banding, and higher harmonics can be superposed without any change to
the solution algorithm.}

\subsection{Coupling between viscoelasticity, damage, and grading}
\label{subsec:coupling}

The coupling between thermo-viscoelasticity, damage evolution, and functional
grading is achieved by introducing the damage variable as a
stiffness-degradation factor in the constitutive response, following the
strain-equivalence principle of continuum damage mechanics
\revb{\cite{lemaitre2005engineering,kachanov1986damage}},
\begin{equation}
  E_\mathrm{eff}(z,T,t) = \bigl(1-D(z,t)\bigr)\,E_{i}(z,T).
  \label{eq:eff_modulus}
\end{equation}
This coupling allows the progressive reduction of stiffness to influence stress
redistribution and damage accumulation during loading, providing a consistent
framework for predicting localized degradation in graded abradable coatings.

\section{Numerical Methodology \revb{and Verification}}
\label{sec:numerical_methodology}

\subsection{Spatial and temporal discretization \revb{and refinement studies}}
\label{subsec:disc}

\revb{The spatial discretization is one-dimensional through the thickness: the
coating column is partitioned into $M$ homogenized subdomains, each treated as a
stratum with piecewise-constant properties. This is formally equivalent to a mesh
of $M$ linear one-dimensional elements of size $h/M$ with a single integration
point per element. No two- or three-dimensional finite-element mesh is employed
and no contact elements exist, so element type and total mesh count in the usual
finite-element sense do not apply; the resolution of the model is defined
entirely by $M$ and $\Delta t$. This subdomain-based semi-analytical discretization is
adopted deliberately because it makes an $N$-realization Monte Carlo loop
computationally tractable while resolving the through-thickness gradients that
dominate this configuration.}

\begin{figure}[!t]
\centering
\includegraphics[width=\textwidth,keepaspectratio]{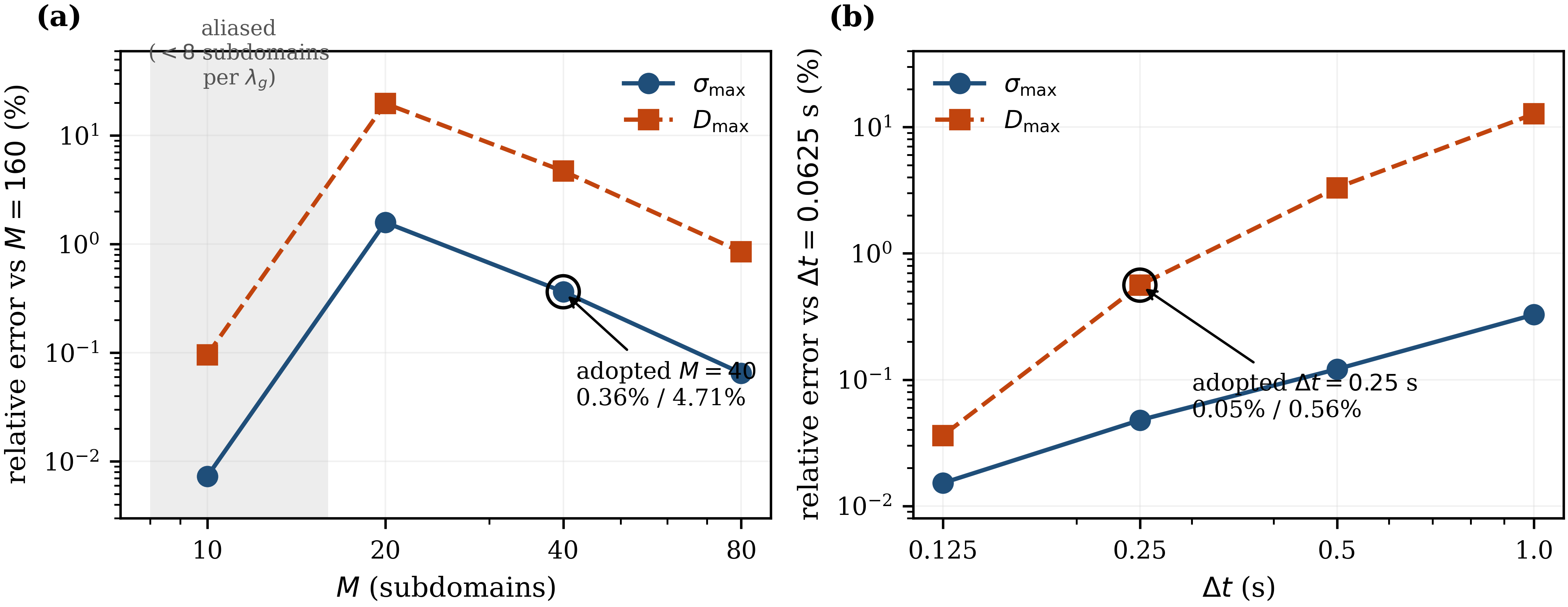}
\caption{\revb{Quantitative spatial and temporal refinement of the reduced-order
solver. (a)~Spatial convergence against the $M=160$ reference is close to second
order; the adopted $M=40$ (circled) gives $0.36\%$ error in maximum stress and
$4.71\%$ in maximum damage. (b)~Temporal convergence against the
$\Delta t=0.0625$~s reference is first order, as expected from the explicit
damage update; the adopted $\Delta t=0.25$~s (circled) gives $0.05\%$ and
$0.56\%$ respectively. The shaded band in~(a) marks resolutions below eight
subdomains per modulation wavelength, where the sinusoidal term of
Eq.~\eqref{eq:fgm_modulus} is aliased and the error is not a monotone function
of $M$.}}
\label{fig:refinement}
\end{figure}

\revb{Both resolutions are established quantitatively against finer declared
references rather than asserted. The spatial series $M=10,20,40,80$ is compared
with the reference $M=160$, and the temporal series
$\Delta t=1.0,0.5,0.25,0.125$~s with the reference $\Delta t=0.0625$~s. The
adopted values are $M=40$, for which the relative errors are $0.36\%$ in maximum
stress and $4.71\%$ in maximum damage, and $\Delta t=0.25$~s, for which the
relative errors are $0.05\%$ and $0.56\%$ respectively. These figures are
collected in Table~\ref{tab:refine} and plotted in Fig.~\ref{fig:refinement}.
The observed behaviour is consistent with the formulation: spatial convergence is
close to second order, while temporal convergence is first order, as expected
from the explicit damage update of Eq.~\eqref{eq:damage_update}. The refinement series is
reported as computed. Below eight subdomains per modulation wavelength the
sinusoidal term of Eq.~\eqref{eq:fgm_modulus} is aliased, so the spatial error is
not a monotone function of $M$ in that range; from $M=20$ upwards it decreases by
a factor of $4.4$ and then $5.6$ per refinement level, consistent with
second-order behaviour, and the temporal error decreases monotonically over the
whole series. What the study establishes is the adopted-point accuracy: at
$\Delta t=1$~s the error in maximum damage is $12.7\%$, which is why that step was
refined, whereas at the adopted $\Delta t=0.25$~s it is $0.56\%$.
The spatial resolution is not selected by refinement alone, and we state the
reason because it is the reverse of the usual argument. Refining $M$ reduces the
discretization error monotonically above the aliasing range, but it also shrinks
the subdomain thickness $h/M$, which must remain at or above the microstructural
feature size for the homogenized description of Section~\ref{subsec:damage} to
hold. At $M=40$, $h/M=50~\mu$m sits just above the coarse end of that range,
whereas $M=80$ would give $h/M=25~\mu$m and place the numerical cell below it.
$M=40$ is therefore adopted as the finest resolution compatible with the
homogenization assumption, and the residual errors it carries relative to the
$M=160$ reference, $0.36\%$ in $\sigma_{\max}$ and $4.71\%$ in $D_{\max}$, are
reported directly rather than described as converged. The adopted step
also samples the fastest retained relaxation mode adequately, since
$\Delta t=\tau_1/20$ with $\tau_1=5$~s, and the recursive convolution update is
exact for piecewise-linear strain within each step.}

\revb{Four closed-form unit tests are used to verify the implementation
independently of any refinement argument, and are summarized in
Table~\ref{tab:unittests}. They isolate, in turn, the viscoelastic kernel, the
free thermal eigenstrain, the constrained expansion mismatch, and the damage
kinetics. All four agree with the analytical solution to within the tolerances
listed. It is emphasized that these are
{verification} tests, establishing that the code solves the intended
equations; they establish nothing about whether those equations describe a
particular real coating.}

\begin{table}[!t]
\centering
\caption{\revb{Spatial and temporal refinement of the reduced-order
through-thickness solver.}}
\label{tab:refine}
\footnotesize
\renewcommand{\arraystretch}{1.15}
\begin{tabular}{@{}p{\dimexpr 0.2535\linewidth-2\tabcolsep\relax}p{\dimexpr 0.2254\linewidth-2\tabcolsep\relax}p{\dimexpr 0.1831\linewidth-2\tabcolsep\relax}p{\dimexpr 0.1690\linewidth-2\tabcolsep\relax}p{\dimexpr 0.1690\linewidth-2\tabcolsep\relax}@{}}
\toprule
\textbf{Refinement item} & \textbf{Series} & \textbf{Reference} &
\textbf{Adopted} & \textbf{Relative error} \\
\midrule
Spatial discretization $M$ & $10,\,20,\,40,\,80$ & $160$ & $40$ &
$\sigma_{\max}$: $0.36\%$ \newline $D_{\max}$: $4.71\%$ \\
Time increment $\Delta t$ (s) & $1.0,\,0.5,\,0.25,\,0.125$ & $0.0625$ & $0.25$ &
$\sigma_{\max}$: $0.05\%$ \newline $D_{\max}$: $0.56\%$ \\
\bottomrule
\end{tabular}
\end{table}

\begin{table}[!t]
\centering
\caption{\revb{Closed-form unit tests used to verify the implementation.}}
\label{tab:unittests}
\footnotesize
\renewcommand{\arraystretch}{1.15}
\begin{tabular}{@{}p{\dimexpr 0.2429\linewidth-2\tabcolsep\relax}p{\dimexpr 0.3286\linewidth-2\tabcolsep\relax}p{\dimexpr 0.2857\linewidth-2\tabcolsep\relax}p{\dimexpr 0.1429\linewidth-2\tabcolsep\relax}@{}}
\toprule
\textbf{Test} & \textbf{Configuration} & \textbf{Analytical reference} &
\textbf{Agreement} \\
\midrule
Relaxation kernel & Step strain, isothermal, $D\equiv0$ &
Eq.~\eqref{eq:prony} evaluated directly & $<10^{-10}$ \\
Free thermal expansion & $\alpha_c=\alpha_s$,
$\varepsilon^{\mathrm{app}}=0$ & Zero stress at all times & $<10^{-12}$ \\
Constrained mismatch & $\Delta\alpha\ne0$, elastic limit
($\tau_m\rightarrow\infty$) & $\sigma=-E\,\Delta\alpha\,\Delta T$ & $<10^{-8}$ \\
Damage kinetics & Constant overstress, $n_d=0$ &
Analytical integration of Eq.~\eqref{eq:damage_law} & $<10^{-6}$ \\
\bottomrule
\end{tabular}
\end{table}

\subsection{Incremental solution algorithm}
\label{subsec:algo}

\begin{figure}[!t]
\centering
\begin{tikzpicture}[
  node distance=6mm and 10mm,
  box/.style={draw, rounded corners=2pt, align=center, inner sep=3pt,
              minimum width=52mm, minimum height=7mm, font=\footnotesize},
  decision/.style={draw, diamond, aspect=2, align=center, inner sep=2pt,
                   minimum width=22mm, minimum height=10mm, font=\footnotesize},
  arr/.style={-{Latex}, line width=0.5pt}
]
\node[box] (A) {Start};
\node[box, below=of A] (B) {Parameter initialization};
\node[box, below=of B] (C) {Set $D_i(t_0)=0$ for all subdomains};
\node[decision, below=of C] (D) {$t < t_\mathrm{final}$?};
\node[box, right=22mm of D] (E) {Update $E_i(t_k)$ (from $T$ and position)};
\node[box, below=of E]  (E2) {\textcolor{black}{Strain split: Eqs.~\eqref{eq:strainsplit}--\eqref{eq:mechstrain}}};
\node[box, below=of E2] (F) {Compute $\sigma_i(t_k)$ (recursive convolution)};
\node[box, below=of F]  (G) {Update $D_i(t_k)$ (Eq.~\eqref{eq:damage_update})};
\node[box, below=of G]  (H) {Advance time and index};
\node[box, below=20mm of D] (J) {End};
\draw[arr] (A) -- (B);
\draw[arr] (B) -- (C);
\draw[arr] (C) -- (D);
\draw[arr] (D.east) -- node[above,font=\scriptsize]{Yes} (E.west);
\draw[arr] (E) -- (E2);
\draw[arr] (E2) -- (F);
\draw[arr] (F) -- (G);
\draw[arr] (G) -- (H);
\draw[arr] (H.west) -- ++(-42mm,0) |- (D.west);
\draw[arr] (D) -- node[left,font=\scriptsize]{No} (J);
\end{tikzpicture}
\caption{Schematic flow diagram describing the iterative resolution algorithm.}
\label{fig:iterative_flowchart}
\end{figure}
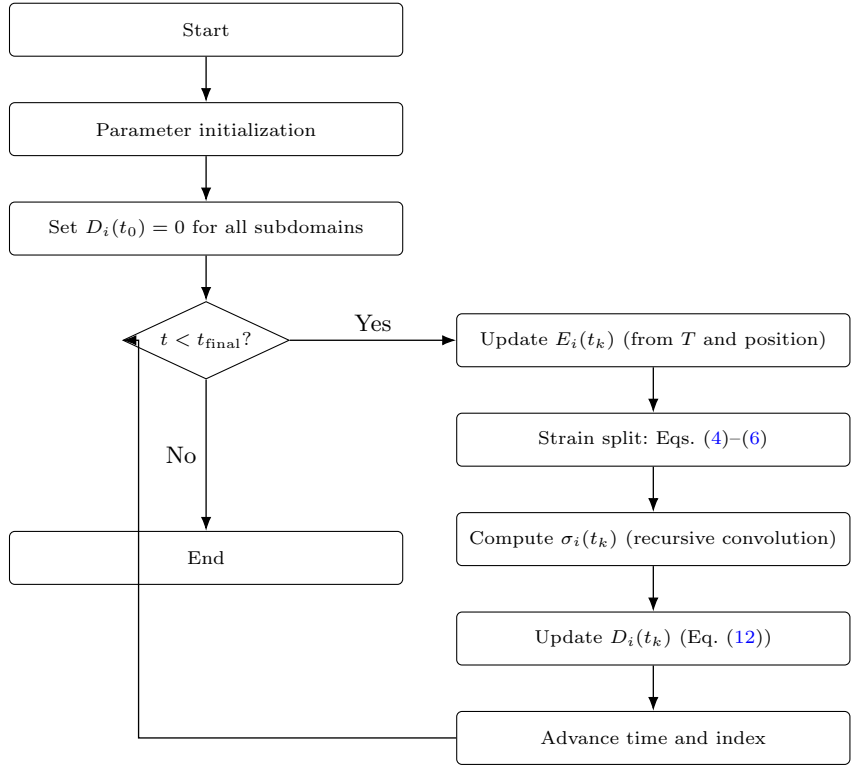

The governing equations are solved using an explicit incremental
time-integration scheme. At each time step $t_k$, the mechanical response and
damage state are updated sequentially for each \revb{subdomain}. The full loop is
summarized in Fig.~\ref{fig:iterative_flowchart} and the corresponding
pseudocode is given in Fig.~\ref{fig:iterative_algorithm_pseudocode}.

\paragraph{Initialization:}
The damage field is initialized as $D_i(t_0)=0$ for all \revb{subdomains}. Initial
temperature, boundary conditions, and material parameters are prescribed.

\paragraph{Update of local viscoelastic properties:}
For each \revb{subdomain} $i$, the graded and periodically modulated elastic
modulus $E_i(z,T)$ is updated according to the local temperature and
through-thickness position. The stiffness degradation induced by damage is
applied through the effective modulus
$E_{\mathrm{eff},i}(t_k)=(1-D_i(t_{k-1}))\,E_i(z,T)$.

\paragraph{\revb{Strain decomposition:}}
\revb{The thermal eigenstrain and the expansion-mismatch strain are evaluated
from the current temperature through Eqs.~\eqref{eq:strainsplit}--\eqref{eq:mismatch},
and subtracted from the applied history to give the mechanical strain increment
of Eq.~\eqref{eq:mechstrain}.}

\paragraph{Stress computation:}
The viscoelastic stress $\sigma_i(t_k)$ is evaluated using a recursive
convolution algorithm based on the Prony-series formulation and the
\revb{mechanical} strain history.

\paragraph{Damage update:}
The damage variable is updated explicitly using the discretized kinetic law
\begin{equation}
  D_i(t_k) = D_i(t_{k-1})
  + A\!\left\langle
            \frac{\revb{\sigma_{\mathrm{eq},i}}(t_k)}{\sigma_\mathrm{crit}}-1
    \right\rangle^{m_d}
    \bigl(1-D_i(t_{k-1})\bigr)^{n_d}\Delta t,
  \label{eq:damage_update}
\end{equation}
with the constraint $0\le D_i\le1$ enforced to preserve physical admissibility.

\paragraph{Time advancement:}
The simulation proceeds to the next time step until the end of the prescribed
loading history.

This explicit incremental strategy provides a robust and computationally
efficient solution framework, particularly well suited for large ensembles of
simulations required by the probabilistic analysis.

\begin{figure}[!t]
\centering
\fbox{%
\begin{minipage}{0.95\linewidth}
\footnotesize\ttfamily
\# Initialization\\
for i in range(1, M+1):\hfill\# M = number of subdomains\\
\hspace*{1.6em}D[i][0] = 0\hfill\# zero initial damage\\
t = delta\_t;\ k = 1\hfill\# index 0 holds the initial condition\\[0.5em]
\# Time loop\\
while t <= t\_final:\\
\hspace*{1.6em}\# Update local property for each subdomain\\
\hspace*{1.6em}for i in range(1, M+1):\\
\hspace*{3.2em}E[i][k] = UpdateElasticModulus(z=z\_i, T=T(i,t))\\[0.3em]
\hspace*{1.6em}\textcolor{black}{\# Thermal eigenstrain and mismatch (Eqs. 4-5)}\\
\hspace*{1.6em}\textcolor{black}{for i in range(1, M+1):}\\
\hspace*{3.2em}\textcolor{black}{eps\_mech[i][k] = ApplStrain(t) - Mismatch(i, T(i,t))}\\[0.3em]
\hspace*{1.6em}\# Viscoelastic recursive convolution\\
\hspace*{1.6em}for i in range(1, M+1):\\
\hspace*{3.2em}sigma[i][k] = ComputeViscoelasticStress(E, eps\_mech, k, i)\\[0.3em]
\hspace*{1.6em}\# Damage update\\
\hspace*{1.6em}for i in range(1, M+1):\\
\hspace*{3.2em}sigma\_eq[i][k] = abs(sigma[i][k])\\[0.3em]
\hspace*{3.2em}rate = DamageRate(sigma\_eq[i][k], D[i][k-1])\\
\hspace*{3.2em}D[i][k] = min(D[i][k-1] + rate * delta\_t, 1.0)\\[0.3em]
\hspace*{1.6em}t = t + delta\_t;\ k = k + 1
\end{minipage}}
\caption{Iterative algorithm pseudocode for the incremental
thermo-viscoelastic--damage solver.}
\label{fig:iterative_algorithm_pseudocode}
\end{figure}

\subsection{Calibration of damage parameters}
\label{subsec:calib}

\revb{The damage law of Eq.~\eqref{eq:damage_law} is phenomenological: the scalar
variable $D$ is an effective stiffness-loss measure, not a physical
representation of the underlying microstructural crack population. The
calibration described below identifies an effective parameter set that reproduces
macroscopic degradation observables; it does not establish a complete
microstructure-to-damage link.}

\revb{The reference dataset is the experimental study of Bertuol on thermally
sprayed AlSi-based abradable coatings~\cite{bertuol2025alsi}, whose
elevated-temperature assessments were conducted at $300\,^\circ$C. That work is
retained as the experimental anchor for the temperature-dependent response and
abradability trends of this coating family: it constrains the expected tendency of
the response within the experimentally documented range and prevents the damage
parameters from being selected without physical guidance. The target observable taken from the reported
stiffness-degradation histories is a normalized stiffness loss of $D=0.45$ after
$60$~s. With $\sigma_\mathrm{crit}=80$~MPa fixed from the onset of measurable
degradation in the same dataset, the triplet $(A,m_d,n_d)$ is identified by
least-squares minimization of the misfit against that degradation curve, yielding
$A=0.12$~s$^{-1}$, $m_d=2.1$ and $n_d=3.2$. \revb{The constant equivalent stress
that reproduces the same target under the identified parameter set follows by
inverting Eq.~\eqref{eq:damage_closed} and is $115$~MPa. We state explicitly what
this number is and is not: it is an internal-consistency benchmark derived from
the identified parameters, not an independently measured stress extracted from
the reference dataset, and it is used only so that the calibration can be checked
in closed form by the reader.} The identified exponents are effective values for the
saturating form of Eq.~\eqref{eq:damage_law} and are reported as such; because
that form carries the damage factor with a positive exponent, they are not
directly comparable with exponents tabulated for the negative-exponent
creep-rupture laws of the same family~\cite{lemaitre2005engineering}.}

\revb{The calibrated set is internally consistent, and this has been verified
analytically rather than asserted. For a constant overstress the damage law
of Eq.~\eqref{eq:damage_law} integrates in closed form to}
\begin{equation}
  \revb{D(t)=1-\Bigl[1+(n_d-1)\,A\,
  \bigl\langle \sigma_{\mathrm{eq}}/\sigma_\mathrm{crit}-1\bigr\rangle^{m_d}\,t
  \Bigr]^{-1/(n_d-1)} .}
  \label{eq:damage_closed}
\end{equation}
\revb{Substituting $\sigma_{\mathrm{eq}}=115$~MPa, $\sigma_\mathrm{crit}=80$~MPa,
$A=0.12$~s$^{-1}$, $m_d=2.1$, $n_d=3.2$ and $t=60$~s gives an overstress of
$0.4375$, a damage rate of $2.116\times10^{-2}$~s$^{-1}$ and $D(60)=0.454$,
which reproduces the calibration target $D=0.45$ to within $1\%$.
Equation~\eqref{eq:damage_closed} is also the reference solution of the
damage-kinetics unit test of Table~\ref{tab:unittests}.}

\revb{The distinction between the experimentally informed range and the range
explored by the benchmark is now stated explicitly. The response up to
$300\,^\circ$C is experimentally informed by the reference dataset. The extension
of the benchmark cycle to $400\,^\circ$C is a constitutive model-based
extrapolation: the temperature dependence over that interval is carried by
the thermo-viscoelastic side of the model through $E_\infty(T)$, $E_m(T)$ and
$\tau_m$, while the damage-kinetics parameters
$(A,m_d,n_d,\sigma_\mathrm{crit})$ are assumed temperature-independent. The
$400\,^\circ$C case is therefore evaluated as a severe numerical benchmark and is
{not} claimed to reproduce the reference protocol or to constitute
experimental validation at that temperature. Because thermally activated
mechanisms could accelerate damage kinetics, the predictions at the dwell
temperature may underestimate degradation and should be interpreted as model-based
extrapolations rather than validated quantitative predictions.
A sensitivity analysis in which $A$ was perturbed by $\pm20\%$ changed the
end-of-cycle damage {at the calibration point} by less than $10\%$
($+7.3\%$ for $A=0.144$~s$^{-1}$ and $-9.0\%$ for $A=0.096$~s$^{-1}$, evaluated
exactly with Eq.~\eqref{eq:damage_closed}). Because $A$ enters
Eq.~\eqref{eq:damage_closed} monotonically, such a perturbation reorders no
realization, so the {ranking} of the ensemble is invariant by construction
and is not itself evidence of robustness; the quantity that does move is the
exceedance probability, since the fraction of realizations above the fixed level
$D_\mathrm{crit}$ shifts with the whole distribution. That probability was
therefore recomputed over the full ensemble under both perturbations: it moves
from $P_e=0.076$ at the nominal $A$ to $0.101$ at
$A=0.144$~s$^{-1}$ and $0.056$ at $A=0.096$~s$^{-1}$. This
calibration-sensitivity range is reported separately from, and must not be
confused with, the sampling interval of Section~\ref{subsec:prob}, which
quantifies Monte Carlo error at fixed $A$. Dedicated cyclic tests at
$400\,^\circ$C would be required for material-specific validation and are
identified as a priority of the experimental programme.}

\subsection{Probabilistic framework \revb{and surrogate cross-check}}
\label{subsec:prob}

Manufacturing and assembly tolerances are modeled as statistically independent
Gaussian random variables centered on nominal geometric dimensions. These
uncertainties are propagated through the numerical model using Monte Carlo
simulations. \revb{The sampled quantities are the manufacturing- and
assembly-related geometric deviations only,
$\delta u_j\sim\mathcal{N}(0,\sigma_{u_j}^2)$ with $\sigma_{u_j}=0.02$~mm. The
subdomain moduli and relaxation times are not sampled independently: they
remain deterministic functions $E_i(z,T)$ and $\tau_m$. The propagation chain
is therefore
$\delta\mathbf{u}\rightarrow\varepsilon^{\mathrm{mech}}\rightarrow\sigma
\rightarrow D\rightarrow$ statistics.}

\revb{The first link of that chain is stated explicitly. The random vector is
$\delta\mathbf{u}=(\delta u_1,\delta u_2,\delta u_3)$, whose three components are
the coating-surface radial-position deviation arising from the manufacturing
thickness tolerance, the counter-face radial-position deviation and the assembly
concentricity deviation. The first component contributes to closure of the
running clearance but does not modify the coating thickness $h$ used in the
constitutive property field of Eq.~\eqref{eq:fgm_modulus}, which is held at its
nominal value in every realization together with the subdomain coordinates and
the modulation profile. They are modelled as independent, zero-mean
and identically distributed, $\delta u_j\sim\mathcal{N}(0,\sigma_u^2)$ with
$\sigma_u=0.02$~mm, and they enter with equal weight because each contributes
directly and additively to the closure of the running clearance. Their sum
perturbs the amplitude of the prescribed approach of Eq.~\eqref{eq:appstrain}
through an effective compliance length $L_{\mathrm{eff}}$,}
\begin{equation}
  \revb{\varepsilon^{\mathrm{app}}_{0}(\delta\mathbf{u})
  =\varepsilon^{\mathrm{app}}_{0}
  +\frac{1}{L_{\mathrm{eff}}}\sum_j \delta u_j ,
  \qquad \delta u_j\sim\mathcal{N}(0,\sigma_{u_j}^{2}),}
  \label{eq:tolmap}
\end{equation}
\revb{so that a symmetric geometric input enters the solver as a symmetric
perturbation of the prescribed strain amplitude and is rendered asymmetric only
by the threshold activation of Eq.~\eqref{eq:damage_law}. Equation~\eqref{eq:tolmap}
is the complete uncertainty map of this study: the through-thickness property
field $E_i(z,T)$, the modulation wavelength and phase, the coating thickness used
in the property law and the relaxation spectrum are identical in every
realization, and only $\varepsilon^{\mathrm{app}}_0$ varies between realizations.
The value of
$L_{\mathrm{eff}}$ is reported in Table~\ref{tab:param_summary}; it is a declared
benchmark quantity of the reduced-order model and not the coating thickness,
since the compliance that absorbs a radial deviation in the physical stage is
that of the whole seal assembly rather than of the coating column alone.}

\revb{Standard Monte Carlo is retained as the reference estimator in
preference to a spectral surrogate such as Polynomial Chaos
Expansion~\cite{sudret2008pce}, for three reasons. First, the response
functionals of interest contain a threshold activation: the Macaulay bracket
in Eq.~\eqref{eq:damage_law} switches subdomains between non-damaging and
damaging regimes, so the response has a kink in the random inputs even though it
is smooth on either side of it. A global polynomial approximation is least
accurate near that kink, which is also where the exceedance mass lies. Second, the targeted quantities
include tail measures, namely $95$th percentiles and the exceedance probability of
$D_\mathrm{crit}$, for which plain Monte Carlo gives unbiased estimates with
quantifiable sampling error, whereas global polynomial surrogates are least
accurate precisely in the tails. Third, the reduced one-dimensional core is
inexpensive and the realizations are embarrassingly parallel, so the full-model
cost did not justify surrogate substitution. This is presented as a methodological
choice among several valid uncertainty-propagation approaches, not as a claim that
PCE is inapplicable.}

\begin{figure}[!t]
\centering
\includegraphics[width=\textwidth,keepaspectratio]{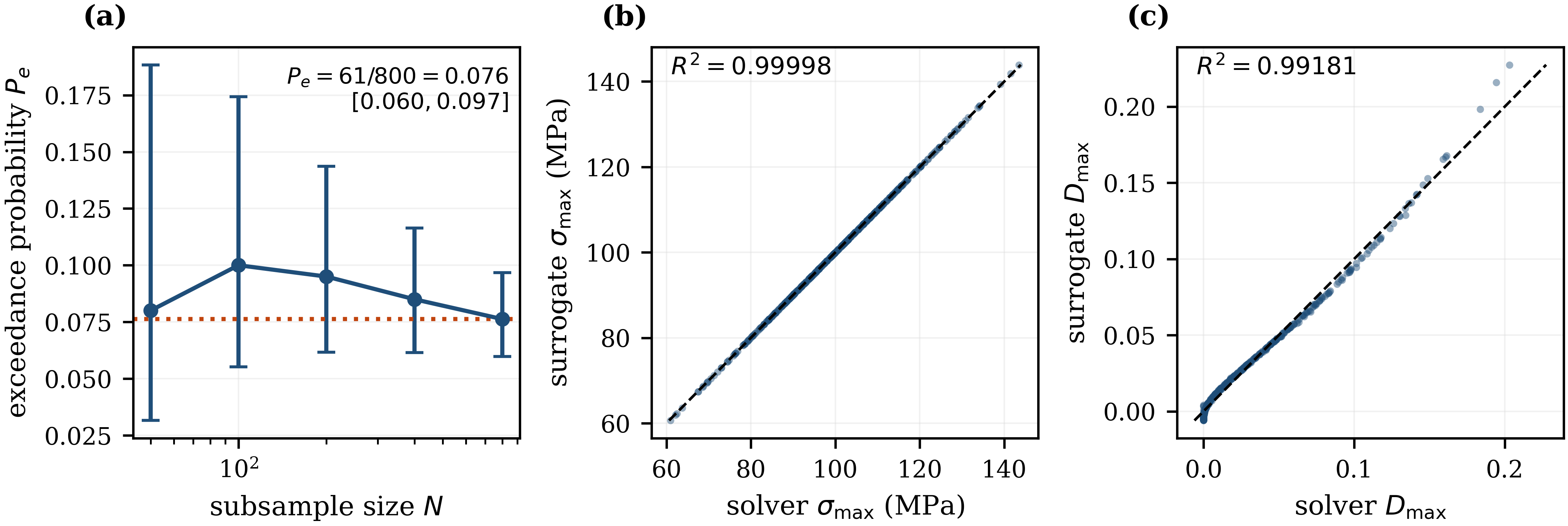}
\caption{\revb{Uncertainty propagation and surrogate verification.
(a)~Exceedance-probability estimate against nested subsample size, with $95\%$
Wilson intervals recomputed from the exceedance counts $4/50$, $10/100$, $19/200$, $34/400$ and $61/800$;
the estimate stabilizes for $N\gtrsim400$ and reaches
$P_e=61/800=0.076$, $[0.060,0.097]$, at $N=800$. (b,c)~Third-order
non-intrusive polynomial response surface fitted on a training subset of $400$
realizations and evaluated on the disjoint hold-out subset of $400$
realizations drawn from the same reference ensemble, giving
$R^2=0.99999$ for maximum stress and $R^2=0.99184$ for maximum damage
against the $1\!:\!1$ line (dashed). The markers are the hold-out set itself, so
the quoted $R^2$ can be recomputed from them. The fit is close over the bulk of
the response and degrades in the upper tail, so the surrogate is used for
screening and continuous-output approximation while the exceedance probability is
estimated by Monte Carlo.}}
\label{fig:mcpce}
\end{figure}

\revb{Sampling precision is assessed by nested subsample sizes rather than by a
single asserted convergence threshold. The reference
ensemble comprises $N=800$ realizations. Fig.~\ref{fig:mcpce}a shows the
exceedance-probability estimate with $95\%$ Wilson confidence intervals as a
function of subsample size; the estimate stabilizes for $N\gtrsim400$, and at
$N=800$ the estimate is $P_e=0.076$ with interval $[0.060,0.097]$.}

\revb{For traceability, the exceedance counts underlying the nested estimates
are stated explicitly: $4/50$, $10/100$, $19/200$, $34/400$ and $61/800$,
giving $\hat{P}_e=0.080$, $0.100$, $0.095$, $0.085$ and $0.076$. The
corresponding $95\%$ Wilson intervals, obtained from
the Wilson expression}
\begin{equation}
  \revb{\mathrm{CI}_{95}=\frac{\hat{p}+z^{2}/2n\;\pm\;
  z\sqrt{\hat{p}(1-\hat{p})/n+z^{2}/4n^{2}}}{1+z^{2}/n} ,
  \qquad z=1.96,}
  \label{eq:wilson}
\end{equation}
\revb{
are $[0.032,0.188]$, $[0.055,0.174]$, $[0.062,0.144]$, $[0.061,0.116]$ and $[0.060,0.097]$. The reported reference value therefore
corresponds to $61$ exceedances out of $800$ realizations, and the
interval half-width contracts as $N^{-1/2}$, as expected for a plain Monte Carlo
estimator.}

\revb{The parameter sensitivity reported in Fig.~\ref{fig:kinetics}b is computed
separately from the tolerance ensemble, by a deterministic one-at-a-time
central-difference scheme, and its definition is given here so that it can be
reproduced. Each parameter $\theta\in\{T_{\max},\tau_1,n_g\}$ is perturbed by
$\pm10\%$ about its Table~\ref{tab:param_summary} value with all others held
fixed, the response being the through-thickness maximum damage
$D_{\max}=\max_z D(z,t_\mathrm{final})$ of the deterministic modulated
configuration. The raw index and its normalized form are}
\begin{equation}
  \revb{S_\theta=\frac{\bigl|D_{\max}(1.1\theta)-D_{\max}(0.9\theta)\bigr|}
  {0.2\,D_{\max}(\theta)} ,
  \qquad
  \bar S_\theta=\frac{S_\theta}{\sum_{\theta'}S_{\theta'}} ,}
  \label{eq:sens}
\end{equation}
\revb{so that $\bar S_\theta$ is a dimensionless elasticity normalized to sum to
unity over the three parameters. The evaluation requires seven solver runs: one
baseline and two per parameter. The scheme is a local screening measure at the
adopted operating point, not a variance-based global index, and it is reported as
such.}

\revb{To cross-check the estimator, a third-order non-intrusive polynomial
response surface was trained and validated on the same reference ensemble: the
$N=800$ realizations are partitioned once into a training subset of $400$ runs
and a disjoint hold-out subset of $400$ runs, so that the surrogate is never
evaluated on data used to fit it and no additional solver runs are required.
The hold-out validation gives $R^2=0.99999$ for maximum stress and
$R^2=0.99184$ for maximum damage (Fig.~\ref{fig:mcpce}b,c), where $R^2$ is
computed on the hold-out set against the $1\!:\!1$ line. The surrogate reproduces
the reduced-order response closely, which is the expected outcome for a smooth
one-dimensional core driven by a small number of Gaussian inputs, and we report
it as such rather than as an argument against spectral methods. What the
comparison does show is where the two approaches differ: the residuals are
largest in the upper tail, precisely the region that sets $P_e$, so the surrogate
is used here for continuous-output approximation and sensitivity screening while
the exceedance probability and its confidence interval are estimated by Monte
Carlo. For an extension of the core to two or three dimensions, the quality of
this fit indicates that a sparse or multi-element polynomial chaos expansion
would be a viable accelerator, with the present Monte Carlo results as the
reference solution.}

\FloatBarrier
\section{Results and Discussion}
\label{sec:results_discussion}

\subsection{Deterministic response and effect of microstructural modulation}
\label{subsec:det}

Simulations using nominal geometric dimensions and \revb{the benchmark} material
parameters are first performed to isolate the influence of functional grading and
deposition-induced microstructural modulation on the thermo-mechanical response
of the abradable coating.

\begin{figure}[!t]
\centering
\includegraphics[width=\textwidth,keepaspectratio]{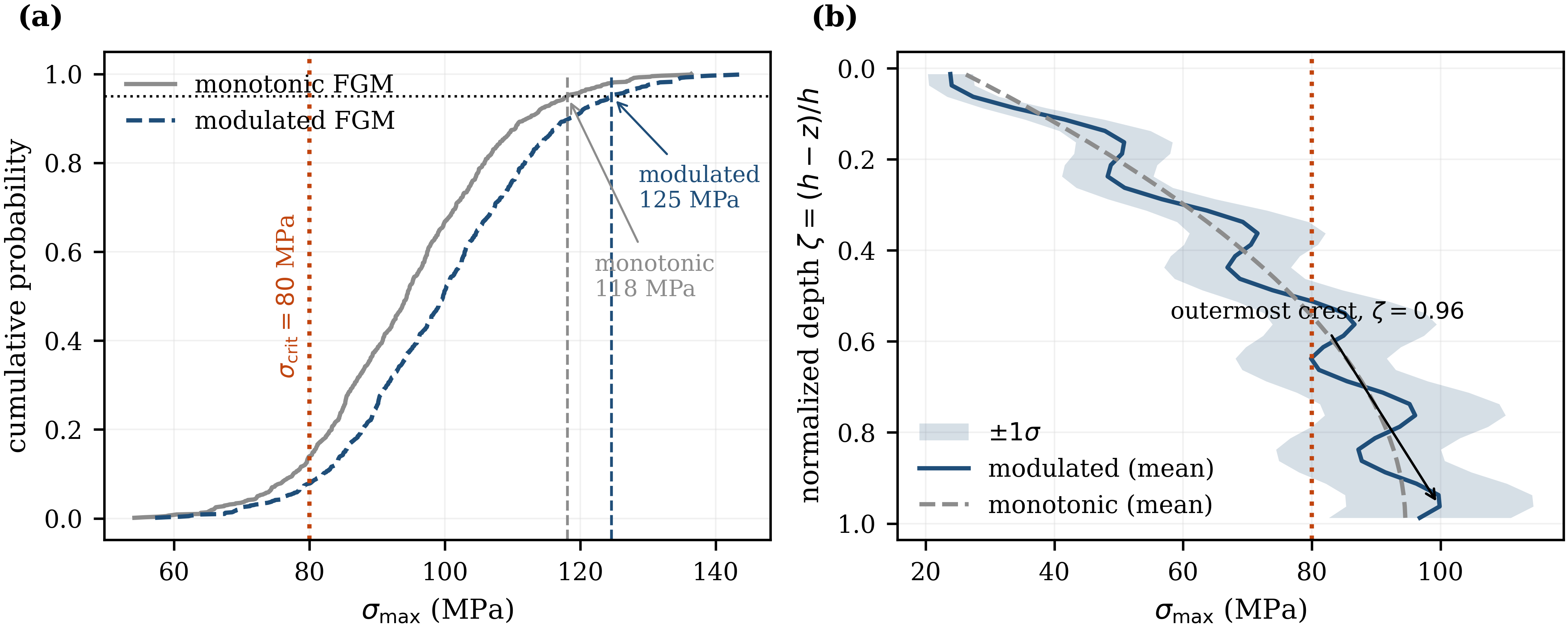}
\caption{Probabilistic stress response. (a)~Cumulative probability distributions
of the maximum stress $\sigma_{\max}$ for the monotonic FGM (solid) and the
periodically modulated FGM (dashed); the dotted vertical line marks the critical
stress threshold, the dotted horizontal line the $95\%$ level, and the two dashed
vertical lines the $95$th-percentile stresses of the two distributions.
(b)~Through-thickness distribution of $\sigma_{\max}$: the solid line is the
ensemble mean of the modulated FGM, the dashed grey line the corresponding
monotonic profile, and the shaded band the $\pm1\sigma$ envelope, all plotted
against the normalized depth below the free surface.}
\label{fig:stress}
\end{figure}

For a monotonic FGM without modulation, the predicted stress field remains
relatively smooth through the coating thickness and does not exhibit pronounced
localized peaks. This behavior is consistent with the stress-mitigating effect of
functionally graded architectures, as reflected by the more moderate upper
quantiles of the maximum-stress statistics compared with the modulated case
(Fig.~\ref{fig:stress}a).

When periodic microstructural modulation is introduced, the stress response
becomes significantly more heterogeneous. Local stiffness fluctuations produce
stress hot spots at specific depths, promoting earlier damage initiation in
selected \revb{subdomains}. This behavior is consistent with the stronger upper
tail and percentile shift observed in the cumulative probability distributions
(Fig.~\ref{fig:stress}a). \revb{The through-thickness profile of
Fig.~\ref{fig:stress}b shows that the maximum equivalent stress {increases}
from the free surface towards the coating--substrate interface. This direction
follows directly from Eqs.~\eqref{eq:mechstrain}, \eqref{eq:convolution}
and~\eqref{eq:fgm_modulus}: with $\alpha_c$ uniform the mechanical strain of
Eq.~\eqref{eq:mechstrain} is essentially uniform through the column, so the
hereditary integral of Eq.~\eqref{eq:convolution} makes the local stress track
the local stiffness, which rises from the compliant dislocator-rich free surface
($E_\mathrm{cer}=12$~GPa at $\zeta=0$) to the metallic-rich interface
($E_\mathrm{met}=48$~GPa at $\zeta=1$). The same ordering is reflected in the
depth-wise damage statistics of Fig.~\ref{fig:damage}b. The modulation
superimposes a
periodic perturbation of wavelength $\lambda_g$, i.e.\ five full periods across
the thickness since $\lambda_g/h=0.2$, on that trend, so that utilisation of the
threshold is highest in discrete bands rather than in one contiguous zone.} The associated damage response is likewise accelerated: the
periodically modulated FGM shows faster mean damage accumulation over time and a
larger separation between the two damage trajectories
(Fig.~\ref{fig:damage}, Fig.~\ref{fig:kinetics}). Overall, these results confirm
that, while the global functional gradient reduces overall stress levels, periodic
modulation governs the local maxima and therefore controls damage localization and
kinetics.

\subsection{Variability induced by geometric tolerances}
\label{subsec:variability}

\begin{figure}[!t]
\centering
\includegraphics[width=\textwidth,keepaspectratio]{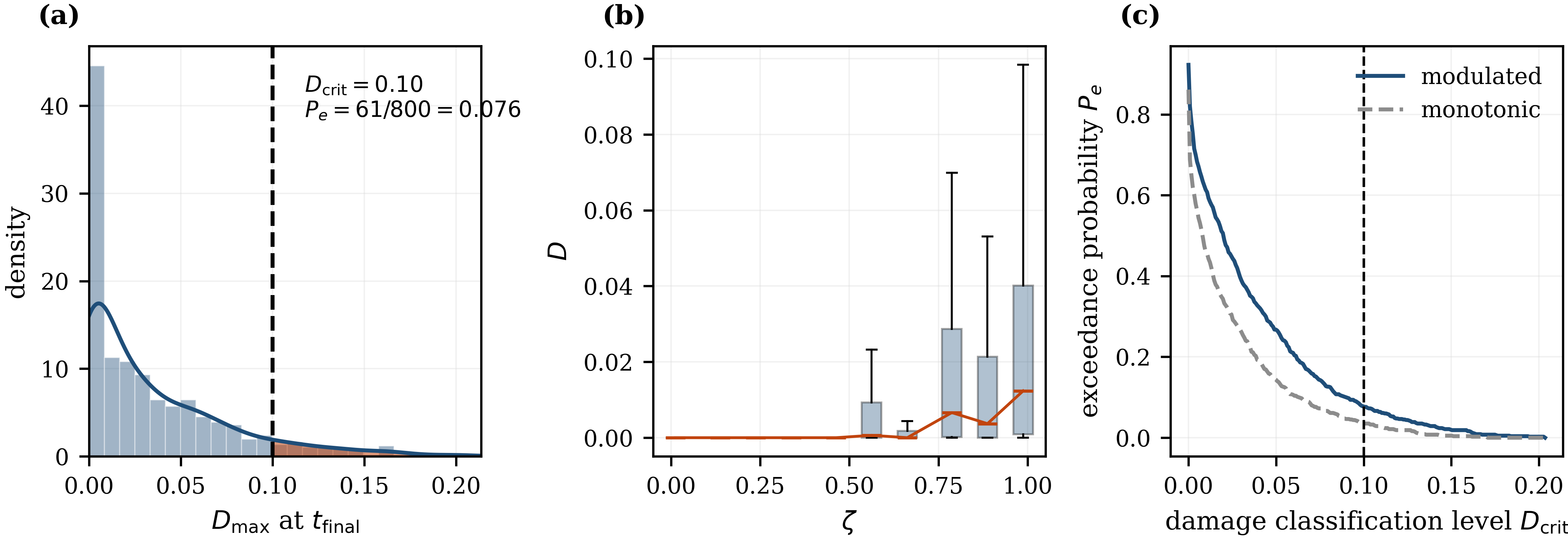}
\caption{Probabilistic damage response ($N=800$ realizations,
\revb{$M=40$, $\Delta t=0.25$~s}).
(a)~Distribution of the \revb{through-thickness maximum} damage
\revb{$D_{\max}=\max_z D(z,t_\mathrm{final})$} at the end of the cycle, with a
kernel density estimate; \revb{the vertical dashed line marks the classification
level $D_\mathrm{crit}=0.10$ and the shaded area is the exceedance tail,
containing $61$ of the $800$ realizations, i.e.\ a fraction $P_e=0.076$}.
(b)~Dispersion of the \revb{local} damage with normalized depth: boxes span the
interquartile range, whiskers the extremes, and the marked line follows the
median. \revb{Panel~(a) aggregates over depth and panel~(b) resolves it, so the
distribution of panel~(a), whose mean is $0.0338$, is shifted above the
local-damage distributions of panel~(b), since each realization contributes its
through-thickness maximum to panel~(a). \revb{(c)~Exceedance probability as a function of the declared
classification level, for both architectures; the dashed line marks the adopted
$D_\mathrm{crit}=0.10$.}}}
\label{fig:damage}
\end{figure}

Introducing geometric tolerances through Monte Carlo simulations produces
substantial dispersion in the mechanical response. \revb{The distribution of the
end-of-cycle damage is shown in Fig.~\ref{fig:damage}a together with a kernel
density estimate and the classification level $D_\mathrm{crit}=0.10$, at which a
realization is classified as exceeding that level. This level is a declared classification
threshold rather than a measured material property, corresponding to a $10\%$
loss of effective stiffness; Fig.~\ref{fig:damage}c reports the exceedance probability as a continuous function of it, so that the sensitivity of the comparison to the selected level can be assessed and the ranking read at any level a designer adopts.} The distribution is strongly
right-skewed. \revb{Its mean is $0.034$ and its median $0.020$, the bulk of the
ensemble lies well below the classification level, and yet
$7.6\%$ of the realizations exceed it. The asymmetry is therefore of
practical rather than cosmetic significance: it is the exceedance mass beyond
$D_\mathrm{crit}$, and not the central tendency, that governs the reliability
metric.}

\revb{The fraction of realizations above the classification level is
$P_e=0.076$, with a $95\%$ Wilson interval of $[0.060,0.097]$ at $N=800$,
against $0.035$ for the monotonic gradient at the same level
(Section~\ref{subsec:prob}).} This confirms that geometric tolerances can produce
rare but severe degradation outcomes that deterministic simulations with nominal
dimensions cannot identify.

Depth-wise variability is further evidenced by the comparative dispersion across
\revb{normalized depth} (Fig.~\ref{fig:damage}b). \revb{Both the median and the
upper quartile increase from the free surface inwards, consistent with the
through-thickness stress profile of Fig.~\ref{fig:stress}b, and the largest
spread and the most extreme values occur in the modulation-controlled bands near
$\zeta\approx0.75$ and $\zeta\approx0.96$, while adjacent depths remain
substantially less damaged. The maximum of the median trend is
reached at the modulation crest closest to the interface rather than at the
interface itself. Eq.~\eqref{eq:fgm_modulus} places the crests of the sinusoidal
term at $z=\lambda_g/4+k\lambda_g$, i.e.\ at
$z=0.1,\,0.5,\,0.9,\,1.3,\,1.7$~mm, which in normalized depth are
$\zeta=0.95,\,0.75,\,0.55,\,0.35,\,0.15$; with $M=40$ the subdomain centred on
the first of these lies at $\zeta=0.9625$. The crest nearest the interface is
therefore $\zeta\simeq0.96$, where the local modulus at the dwell temperature
reaches $32.9$~GPa against $31.6$~GPa at the next crest inboard, $\zeta=0.75$. The
outermost crest is consequently the most heavily loaded band, and it, rather
than the interface node itself, carries the maximum of the median damage.} This depth-dependence has direct practical implications
for non-destructive inspection: \revb{sub-surface} depths \revb{in the outer
modulation crests, i.e.\ from roughly three quarters of the coating thickness
below the free surface down to the interface,} are systematically more
damage-prone and should be prioritized in evaluation protocols.

\subsection{Combined probabilistic effects of grading, modulation, and
tolerances}
\label{subsec:combined}

\begin{figure}[!t]
\centering
\includegraphics[width=\textwidth,keepaspectratio]{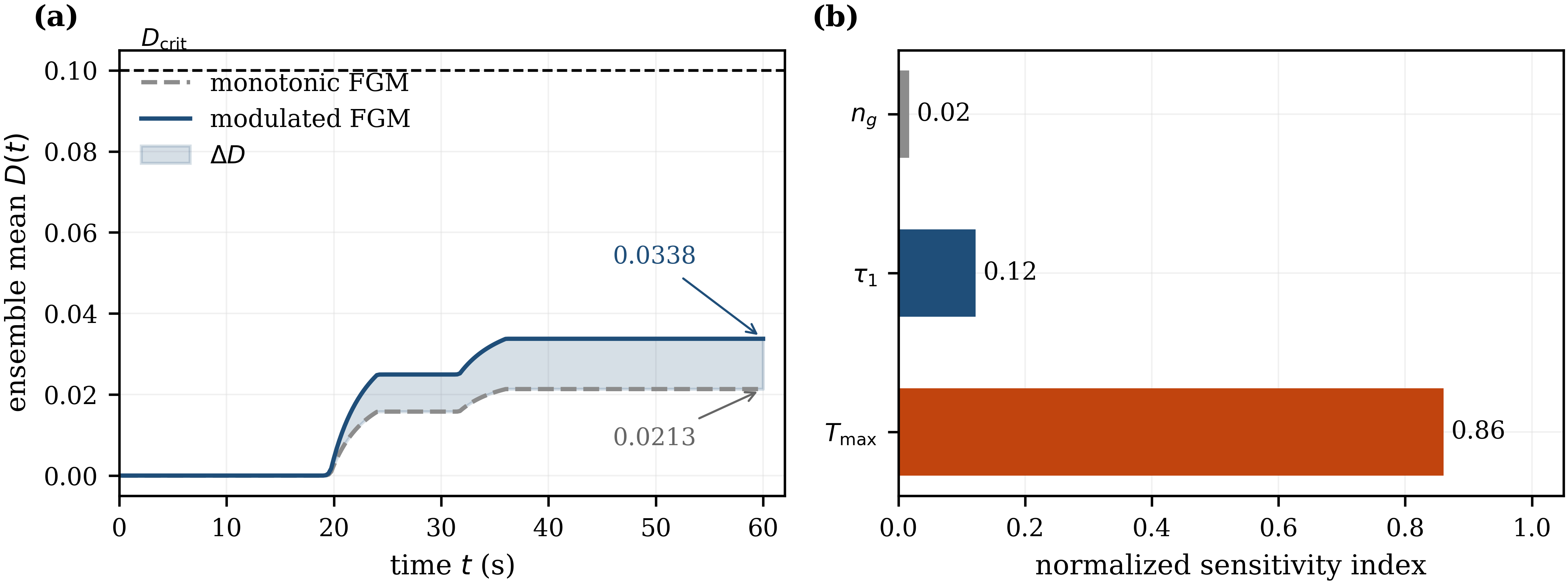}
\caption{(a)~Comparison of the \revb{ensemble} mean damage $D(t)$ for the
monotonic and periodically modulated FGM, with the divergence $\Delta D$ shaded;
\revb{the two curves terminate at $0.02135$ and $0.03376$ respectively, in agreement
with the end-of-cycle distribution of Fig.~\ref{fig:damage}a, and both remain
below the classification level $D_\mathrm{crit}=0.10$, so the exceedance
probability of Section~\ref{subsec:variability} is generated entirely by
tolerance-induced dispersion.} \revb{(b)~Normalized {local} sensitivity indices, computed independently
of the Monte Carlo ensemble of panel~(a) using the seven-run one-at-a-time
central-difference procedure of Eq.~\eqref{eq:sens}, at $M=40$ and
$\Delta t=0.25$~s.}}
\label{fig:kinetics}
\end{figure}

A direct comparison between probabilistic responses for the monotonic FGM and the
periodically modulated FGM highlights the combined influence of microstructural
heterogeneity and geometric variability. Modulation increases both the rate and
the level of damage accumulation relative to the unmodulated case
(Fig.~\ref{fig:kinetics}a). When geometric tolerances are added, response
distributions broaden and exhibit increased dispersion, as shown in
Fig.~\ref{fig:stress}a and Fig.~\ref{fig:damage}a.

\revb{In both cases the mean damage increases monotonically with time, reflecting
progressive degradation under the applied thermo-mechanical loading. However, the
periodically modulated FGM exhibits consistently higher damage at all time
instants. At $t=60$~s, the mean damage reaches $0.03376$ in the modulated case
compared with $0.02135$ in the unmodulated case, an increase of $58\%$, confirming
that modulation can intensify damage kinetics and yield a more severe
end-of-cycle state. \revb{Both values are ensemble means of the end-of-cycle
distributions and therefore coincide with the mean of Fig.~\ref{fig:damage}a.}
Damage begins to accumulate during the heating ramp, as
soon as the expansion mismatch alone drives the equivalent-stress utilisation
$\sigma_{\mathrm{eq}}/\sigma_\mathrm{crit}$ above unity in the stiffer subdomains, and the gap
between the two curves widens progressively through the thermal dwell, because
the modulated case activates in more subdomains and accumulates faster in each.}

Neglecting microstructural modulation therefore leads to systematic
underestimation of (i)~the central tendency of damage growth and (ii)~the
probability of critical exceedance. The depth-wise statistics further indicate
that this interaction is depth-dependent and not spatially uniform.

\revb{Fig.~\ref{fig:kinetics}b reports normalized sensitivity indices for the
parameters that the model actually contains, computed by the one-at-a-time
central-difference scheme defined in Section~\ref{subsec:prob}. No friction-coefficient index
appears, because the implemented formulation contains no friction law
(Table~\ref{tab:scope}); the indices are normalized to sum to unity. The dwell
temperature dominates ($0.86$), through its effect on both the expansion
mismatch of Eq.~\eqref{eq:mismatch} and the temperature-dependent moduli, followed
by the leading relaxation time ($0.12$); the gradient exponent contributes
only marginally ($0.02$) because it reshapes the property profile without
changing the peak stiffness at the interface, where damage localizes. The influence of the geometric tolerance is
quantified separately, and directly, by the ensemble dispersion of
Figs.~\ref{fig:stress} and~\ref{fig:damage} rather than by a single scalar index.}

\subsection{Discussion and implications for coating reliability}
\label{subsec:discussion}

Taken together, the deterministic and probabilistic results show that
functionally graded designs can reduce global stress levels but do not
necessarily guarantee improved durability when deposition-induced heterogeneity
and geometric tolerances are present. The acceleration of damage in the
periodically modulated case and the emergence of increased damage dispersion
under tolerances demonstrate that local stiffness modulation and dimensional
variability can dominate durability outcomes. \revb{The probabilistic results
further show that tolerance-induced dispersion yields an exceedance probability
of $P_e=0.076$ for the declared level $D_\mathrm{crit}=0.10$, more than double the
$0.035$ obtained for the monotonic gradient at the same level.} The observed trends are consistent with
reliability and damage-mechanics perspectives reported in the
literature~\cite{lemaitre2005engineering,haldar2000probability}.

\revb{From a validation standpoint it is important to state precisely what has and
has not been established. What has been established is numerical
verification: the implementation reproduces closed-form solutions for the
viscoelastic kernel, the thermal eigenstrain and the damage kinetics
(Table~\ref{tab:unittests}); the residual discretization errors at the adopted
resolutions are quantified against finer references (Table~\ref{tab:refine}); the sampling estimate is
stable with quantified confidence intervals; and the reference estimator has been
cross-checked against a response surface validated on a disjoint hold-out subset
(Fig.~\ref{fig:mcpce}). What has not been established is experimental
validation. The damage parameters are anchored on published data up to
$300\,^\circ$C, and the extension of the benchmark to $400\,^\circ$C is a
model-based extrapolation (Section~\ref{subsec:calib}). The qualitative trends,
namely the stress-mitigating role of grading and the stress-amplifying effect of
periodic modulation, are independently supported in the
literature~\cite{wang2025ceramic,huang2024review,yildirim2011periodic,%
chen2021modulation}, but the present numbers are benchmark outputs, not
predictions for a specific coating.}

\revb{Pathways for experimental validation of the framework include:
(i)~instrumented thermal-cycling experiments on coupon-scale graded specimens with
characterized microstructural gradients and measured expansion coefficients;
(ii)~through-thickness property identification by nanoindentation or ultrasonic
mapping to measure $\Delta E_g$ and $\lambda_g$ directly; (iii)~cyclic
degradation tests at $400\,^\circ$C to replace the present extrapolation with
measured kinetics; and (iv)~for any future wear-oriented extension, dedicated
rub-rig testing with measured contact geometries, along the lines of the
labyrinth-seal rubbing campaigns reported by
Pychynski~\cite{pychynski2016labyrinth} for honeycomb inlet liners, which would
additionally require an identified tribological law together with explicit
contact and sliding quantities that the present formulation does not contain.}

\FloatBarrier
\section{Implications, Limitations, and Prospects}
\label{sec:implications_limits_prospects}

The proposed framework provides actionable guidance for the design and
maintenance planning of functionally graded abradable coatings in turbine
labyrinth seals. In particular, the results show that deposition-induced
microstructural periodicity and geometric tolerances can control the location and
severity of damage, which motivates their explicit inclusion during coating
design and manufacturing specification. \revb{These implications are stated
within the declared scope: they concern architecture screening and tolerance
allocation for the benchmark system, not service-life prediction for a specific
coating.}

\subsection{Limitations}
\label{subsec:limitations}

The main limitations of the present model are the use of a scalar isotropic damage
variable \revb{(valid only at the REV/subdomain scale established in
Section~\ref{subsec:damage}, and used only for macroscopic stiffness degradation
rather than for crack-path prediction)}, the neglect of irreversible viscoplastic
effects, and the absence of coupled fluid--structure--thermal interactions, each
of which may affect quantitative predictions under high-load or transient regimes.
\revb{The following additional limitations follow from the declared scope and are
stated explicitly:}

\begin{itemize}
\item \revb{\textbf{No contact, friction or wear:} the solver contains no
  rotor--stator contact algorithm, no tangential sliding interface, no Coulomb
  friction law, no frictional heat source and no material-removal kinetics. The
  mechanical excitation enters as a prescribed equivalent normal-strain history.
  Consequently no wear depth, incursion depth or abradability metric is predicted,
  and any such extension would require an independently identified tribological
  law together with explicit contact and sliding quantities.}
\item \revb{\textbf{Porosity is homogenized:} porosity enters only through the
  effective properties of each subdomain and through its contribution to the
  modulation amplitude. Individual pores are never resolved, so pore-scale stress
  concentrations are outside the model. Moreover, because the temperature history
  is prescribed rather than solved from a heat-conduction problem, the
  thermal-barrier role of porosity does not influence the solution; resolving it
  would require solving the coupled conduction problem with a
  porosity-dependent conductivity field.}
\item \revb{\textbf{Constant damage threshold:} $\sigma_\mathrm{crit}$ is an
  effective homogenized value held constant through the thickness
  (Section~\ref{subsec:damage}). This is adequate for the global structural
  response but cannot resolve local microstructural damage initiation; a spatially
  varying threshold $\sigma_\mathrm{crit}(z,T,\phi)$ is the natural refinement.}
\item \revb{\textbf{Calibration range.} The damage kinetics are anchored on data
  at $300\,^\circ$C and assumed temperature-independent up to $400\,^\circ$C, so
  the dwell-temperature predictions are model-based extrapolations that may
  underestimate degradation, rather than validated quantitative predictions.}
\item \revb{\textbf{Benchmark parameter set:} Expansion coefficients, relaxation
  times and modulation parameters are declared benchmark values requiring material
  identification (Table~\ref{tab:param_summary}); the simulated case is not an
  identified model of a specific coating.}
\item \revb{\textbf{Prescribed rather than solved loading:} Both the thermal and
  the mechanical histories are inputs. The framework therefore quantifies how a
  {given} excitation interacts with material heterogeneity and dimensional
  variability; it does not predict the excitation itself.}
\end{itemize}

Future work shall focus on incorporating viscoplasticity and anisotropic damage
formulations, extending the model to transient fluid--structure--thermal
coupling, \revb{introducing a spatially variable damage threshold, and coupling
the column to an explicit contact and material-removal model should wear
prediction become an objective,} and validating predictions experimentally using
coatings with controlled microstructural gradients. The probabilistic component
can also be extended toward stochastic optimization of graded architectures to
minimize exceedance probability.

\subsection{Practical implications for design and maintenance}
\label{subsec:practical}

The framework's outputs map into the decision variables of coating development
and service practice, \revb{within the declared scope,} as detailed below.

\textbf{Manufacturing tolerance specification:} geometric variability is a major
source of uncertainty and can increase the likelihood of exceeding the
classification level. \revb{Using $D_\mathrm{crit}=0.10$, the end-of-cycle
distribution gives $P_e=0.076$ with interval $[0.060,0.097]$
(Fig.~\ref{fig:damage}a).} This motivates tighter tolerance control to reduce the
probability of exceeding the damage classification level. By re-running the Monte Carlo analysis under
progressively tighter tolerance bands and monitoring the resulting reduction in
exceedance probability, engineers can identify the tolerance tightening that
achieves a target reliability level while minimizing manufacturing
cost~\cite{haldar2000probability,sudret2008pce}.

\textbf{Coating design optimization:} because the periodic modulation amplitude
$\Delta E_g$ and wavelength $\lambda_g$ directly govern stress hot-spot
intensity (Section~\ref{subsec:fgm_mod}), deposition process parameters that
control these quantities, such as spray angle, traverse speed and powder
morphology, can be linked to reliability outcomes. Designers can use the
framework to identify modulation amplitudes that keep the $95$th-percentile
stress below a target, or to specify a gradient exponent $n_g$ that minimizes
mean damage at end of cycle~\cite{suresh1998fundamentals,miyamoto1999fgm}.
Explicit inclusion of microstructural modulation in design models avoids
underestimation of local stress hot spots and supports more realistic
optimization of the gradient profile.

\textbf{Inspection prioritization:} probabilistic mapping of high-damage outcomes
supports targeted inspections. The depth-wise damage statistics
(Fig.~\ref{fig:damage}b) identify the depth zones with the highest variance and
most extreme values, suggesting that non-destructive evaluation should be
calibrated to detect damage at \revb{sub-surface depths, around
the modulation-controlled bands nearest the interface, at $\zeta\approx0.75$ and
$\zeta\approx0.96$,} rather than near the surface.
In a parallel effort within physics-informed reliability engineering, our previous
work~\cite{dhibi2026physics} demonstrated that embedding fracture-mechanics priors
into deep learning improves remaining-useful-life predictions for
rolling-element bearings; as a perspective, the multiphysics damage data produced
by the present framework could extend such approaches through data-driven
calibration. \revb{We note explicitly that the present framework does not itself
produce a remaining-useful-life estimate, since that would require the
service-life modelling that is outside the declared scope.}

\subsection{Comparison with the Literature}
\label{subsec:literature}

The trends observed in this study are consistent with published experimental,
analytical, and probabilistic investigations on graded materials, modulated
coatings, and structural reliability. Periodic modulation of material properties
is known to promote local stress amplification in graded and periodically
structured media~\cite{yildirim2011periodic,chen2021modulation}. This is directly
aligned with the present results, where the periodically modulated FGM produces
more localized stress hot spots and faster damage accumulation than the monotonic
FGM. In particular, Chen et al.~\cite{chen2021modulation} demonstrated
experimentally that modulation geometry governs both mechanical and tribological
performance in nano-multilayer coatings. This supports the central interpretation
of the present framework: microstructural modulation should not be treated as a
minor geometric perturbation, because it can reshape the local stress field and
therefore influence damage initiation and growth.

The importance of accounting for dimensional tolerances and uncertainty
propagation in structural performance assessment is well established in classical
reliability references~\cite{haldar2000probability,melchers1999structural}. The
present work extends this reliability perspective to abradable coating analysis by
propagating geometric tolerances through the full thermo-viscoelastic--damage
solver rather than evaluating uncertainty only at the level of simplified output
statistics. \revb{The modulated property field modifies local stiffness and stress
redistribution, while the sampled geometric deviation modifies the equivalent
mechanical input; their combined effect produces response distributions that
cannot be captured by deterministic simulations at nominal geometry.}

\revb{Recent probabilistic fatigue and damage-modelling studies likewise report
non-negligible upper-tail probabilities under cyclic
loading~\cite{li2021stochastic}. Those results provide qualitative support for
examining the response distribution rather than the nominal prediction alone. No
direct quantitative comparison with the $P_e$ reported here is drawn, because the
material systems, the loading conditions and the classification criteria differ,
and because $D_\mathrm{crit}$ is a declared screening level rather than an
experimentally identified failure criterion; the two quantities are therefore not
measures of the same event.}

\revb{Finally, some limitations should be borne in mind when comparing these
results with the broader literature. The absence of irreversible plasticity may
lead to quantitative deviations for coatings that exhibit pronounced viscoplastic
deformation; the lack of coupled fluid--structure--thermal dynamics limits direct
applicability under fast transients; and comparisons with rub-rig or abradability
studies are necessarily qualitative, because the present formulation solves
neither contact nor material removal.}

\FloatBarrier
\section{Conclusions and Perspectives}
\label{sec:conclusions}

This study developed a unified multiphysics--probabilistic framework for
predicting thermo-viscoelastic damage in functionally graded abradable coatings
for turbine labyrinth seals. The model integrates temperature-dependent
viscoelasticity, \revb{thermal eigenstrain and coating--substrate expansion
mismatch,} progressive damage, deposition-induced microstructural modulation, and
geometric tolerances within a single reliability-oriented formulation. The
central novelty lies in their unified treatment within a single computational
pipeline: coupling periodic property modulation with Monte Carlo tolerance
propagation produces nonlinear probabilistic response distributions that decoupled analyses
cannot replicate, and the resulting reliability metrics, exceedance
probabilities and sensitivity indices, are directly usable for tolerance
specification and inspection planning.

The results show that, although functional grading can reduce global stress
levels, deposition-induced microstructural heterogeneity may dominate damage
localization and accelerate degradation at specific depths. The probabilistic
analysis further indicates that geometric tolerances amplify response dispersion
and increase the exceedance probability, highlighting the importance of accounting for
manufacturing variability in reliability assessments. Consequently,
deterministic analyses based on idealized monotonic gradients can underestimate
the likelihood of localized damage and of exceeding the declared classification level.

\revb{The contribution is deliberately positioned as a numerically
verified reduced-order methodology. The solved domain is a local
through-thickness coating column driven by prescribed histories; explicit
rubbing contact, friction, frictional heating, material removal and wear-depth
prediction are outside the formulation and are not reported. The simulated case
is a declared generic benchmark, and numerical verification, established through
closed-form unit tests, spatial and temporal refinement, sampling convergence and
a hold-out-validated response surface, is distinguished throughout from
experimental validation, which remains future work.}

Overall, the proposed framework provides a practical basis for robust coating
design, tolerance control, and inspection planning. While demonstrated for
abradable coatings in turbine labyrinth seals, the framework is general and
applicable to any layered or graded system, including thermal barrier coatings,
environmental barrier coatings, and functionally graded structural components,
where deposition-induced heterogeneity and dimensional variability jointly govern
reliability. Future work shall incorporate viscoplasticity and anisotropic damage,
a spatially variable damage threshold, coupled fluid--structure--thermal
interactions, and experimental validation using graded coating systems with
controlled microstructural features.

\section*{Nomenclature}
\begin{description}\itemsep2pt
\item[$D$] Scalar damage variable ($0\le D\le1$), defined at the REV/subdomain scale.
\item[$D_\mathrm{crit}$] Critical damage classification level.
\item[$E(t,T)$] Temperature-dependent relaxation modulus.
\item[$E_\infty(T)$] Long-term (equilibrium) modulus in the Prony representation.
\item[$E_m(T)$] $m$-th Prony-series modulus coefficient.
\item[$\tau_m$] $m$-th relaxation time, \revb{held constant with temperature in the present benchmark}.
\item[$\sigma(z,t)$] Viscoelastic stress response.
\item[$\sigma_\mathrm{crit}$] Effective critical stress threshold governing damage activation.
\item[$\sigma_{\max}$] Maximum stress statistic, \revb{$\sigma_{\max}=\max_t|\sigma(z,t)|$}.
\item[\revb{$A$}] \revb{Damage rate coefficient (s$^{-1}$).}
\item[\revb{$m_d$}] \revb{Damage exponent acting on the overstress.}
\item[\revb{$n_d$}] \revb{Damage exponent acting on the saturation term.}
\item[$E_i(z,T)$] Graded and modulated modulus in subdomain $i$ at depth $z$.
\item[$E_\mathrm{eff}$] Damage-degraded effective modulus, $E_\mathrm{eff}=(1-D)E_i$.
\item[$z$] Through-thickness coordinate, $z=0$ at the interface, $z=h$ at the free surface.
\item[\revb{$\zeta$}] \revb{Normalized depth below the free surface, $\zeta=(h-z)/h$.}
\item[$h$] Total coating thickness.
\item[$n_g$] Functional gradient exponent.
\item[\revb{$\Delta E_g$}] \revb{Amplitude of periodic microstructural modulation (GPa).}
\item[$\lambda_g$] Wavelength of microstructural modulation.
\item[\revb{$\phi$}] \revb{Homogenized total porosity.}
\item[\revb{$\alpha_c,\alpha_s$}] \revb{Coating and substrate thermal expansion coefficients.}
\item[\revb{$\Delta\alpha$}] \revb{Expansion mismatch, $\Delta\alpha=\alpha_c-\alpha_s$.}
\item[\revb{$\Delta T_{\max}$}] \revb{Peak temperature rise above $T_\mathrm{ref}$ over the benchmark cycle.}
\item[\revb{$T_\mathrm{ref}$}] \revb{Stress-free reference temperature.}
\item[\revb{$\varepsilon^{\mathrm{th}}$}] \revb{Thermal eigenstrain.}
\item[\revb{$\varepsilon^{\mathrm{mis}}$}] \revb{Coating--substrate expansion mismatch strain.}
\item[\revb{$\varepsilon^{\mathrm{mech}}$}] \revb{Mechanical (stress-producing) strain.}
\item[\revb{$\varepsilon^{\mathrm{app}}$}] \revb{Prescribed equivalent applied normal strain.}
\item[$\Delta t$] Time step used in incremental integration.
\item[$t_\mathrm{final}$] Final simulation time.
\item[$M$] Number of through-thickness homogenized subdomains.
\item[$M_p$] Number of Prony terms in the viscoelastic series.
\item[$N$] Number of Monte Carlo realizations.
\item[$\delta\mathbf{u}$] Vector of sampled geometric deviations (tolerances).
\item[\revb{$L_{\mathrm{eff}}$}] \revb{Effective compliance length mapping a geometric deviation onto the prescribed strain amplitude, Eq.~\eqref{eq:tolmap}.}
\item[\revb{$P_e$}] \revb{Exceedance probability of $D_\mathrm{crit}$.}
\item[\revb{$D_{\max}$}] \revb{Through-thickness maximum damage at the end of the cycle.}
\item[\revb{$R^2$}] \revb{Hold-out coefficient of determination of the polynomial response surface.}
\item[\revb{$n_p,t_r,t_d$}] \revb{Number of approach pulses, ramp time and plateau time of the prescribed strain waveform.}
\end{description}

\section*{Declarations}

\subsection*{Conflict of Interest}
The authors declare no competing interests.

\subsection*{Data and Code Availability}
The simulation code developed for this study, including the
thermo-viscoelastic--damage solver described in
Section~\ref{sec:numerical_methodology} and the Monte Carlo automation scripts
used for the probabilistic tolerance analysis in Section~\ref{subsec:prob}, will
be made available upon reasonable request to the corresponding author
(felmellouhi@hbku.edu.qa). The code implements the incremental
thermo-viscoelastic--damage solver, the through-thickness property-field
construction, \revb{the thermal-eigenstrain and expansion-mismatch evaluation,}
the damage-update procedure, and the Monte Carlo propagation of geometric
tolerances. 

\revb{All deterministic and probabilistic simulation inputs required for
reproducibility are reported in the manuscript, including the nominal geometry,
the through-thickness discretization, the prescribed thermal and mechanical
histories, the Prony parameters, the expansion coefficients, the functional
gradient and modulation parameters, the damage parameters, the geometric
tolerance definitions, the Monte Carlo settings, and the post-processing metrics
(Table~\ref{tab:param_summary}). The refinement data of Table~\ref{tab:refine},
the unit-test tolerances of Table~\ref{tab:unittests}, and the surrogate
validation metrics of Fig.~\ref{fig:mcpce} constitute the verification record.
Every quantity reported in more than one place is traceable to
Table~\ref{tab:param_summary} and to the equation that defines it: the peak
mismatch strain of Eq.~\eqref{eq:mismax} to the expansion coefficients
$\alpha_c$ and $\alpha_s$; the exceedance probability $P_e=0.076$ to the count
$61/800$ and the Wilson interval $[0.060,0.097]$; and the calibrated triplet
$(A,m_d,n_d)$ to the closed-form solution of Eq.~\eqref{eq:damage_closed}.
The calibration data used for anchoring the damage-law parameters are derived
from the published study of Bertuol~\cite{bertuol2025alsi} and are therefore not
redistributed independently; the calibration procedure and all parameter values
required to reproduce the simulations are documented in
Section~\ref{subsec:calib}.}

\subsection*{Acknowledgement}
This work was supported by the Knowledge Transfer Program--Qatar (KTP-Q) project
No.\ KTPQ01-0530-240004 in partnership with ADGS Computer Systems, and by the
Mitacs Accelerate program under project IT47288. The statements herein are solely
the responsibility of the authors.


\end{document}